\documentclass[transmag]{IEEEtran}
\usepackage{latexsym}
\usepackage{graphicx}
\usepackage{amsfonts,amssymb,amsmath}
\usepackage{hyperref}
\usepackage{epsfig}
\usepackage{algorithm}
\usepackage{algorithmic}
\usepackage{cite}
\usepackage{caption}
\def\BibTeX{{\rm B\kern-.05em{\sc i\kern-.025em b}\kern-.08em T\kern-.1667em\lower.7ex\hbox{E}\kern-.125emX}}
\begin{document}

\newtheorem{Thm}{Theorem}[section]
\newtheorem{Cor}[Thm]{Corollary}
\newtheorem{Lem}[Thm]{Lemma}
\newtheorem{Prop}[Thm]{Proposition}
\newtheorem{Def}[Thm]{Definition}
\newtheorem{Rem}[Thm]{Remark}
\newtheorem{Exa}[Thm]{Example}
\newtheorem{Assump}[Thm]{Assumption}

 \title{Natural Thresholding Algorithms for Signal Recovery with Sparsity}

\author{Yun-Bin Zhao, \IEEEmembership{Member, IEEE} and Zhi-Quan Luo, \IEEEmembership{Fellow, IEEE}
\thanks{This work was supported by the National Natural Science Foundation of China
(NSFC) under grants \#12071307 and \#61571384. Part of the work was done while the first author worked at the School of Mathematics, University  of Birmingham, Birmingham B15 2TT,  United Kingdom.}
\thanks{Yun-Bin Zhao is with the Shenzhen Research Institute of Big Data, Chinese University of Hong Kong, Shenzhen, Guangdong Province, China (Email: yunbinzhao@cuhk.edu.cn).}
\thanks{Zhi-Quan Luo is with the Shenzhen Research Institute of Big Data, Chinese University of Hong Kong, Shenzhen, Guangdong Province, China (Email: luozq@cuhk.edu.cn).}}

\IEEEtitleabstractindextext{
\begin{abstract}
The algorithms based on the technique of optimal $k$-thresholding (OT) were recently proposed for signal recovery, and they are very different from the traditional family of hard thresholding methods. However, the computational cost for OT-based algorithms remains high at the current stage of their development. This
  stimulates the development of the so-called natural thresholding (NT) algorithm and its variants in this paper. The family of NT algorithms is developed through the first-order approximation of the so-called regularized optimal $k$-thresholding model, and thus the computational cost for this family of algorithms is significantly lower than that of the OT-based algorithms. The guaranteed performance of NT-type algorithms for signal recovery from noisy measurements is shown under the restricted isometry property and concavity of the objective function of regularized optimal $k$-thresholding model.  Empirical results indicate that the NT-type  algorithms  are robust and very comparable to several mainstream algorithms for sparse signal recovery.
\end{abstract}

 }


\maketitle

\section{Introduction}

\IEEEPARstart{U}{nder} the sparsity assumption, the problem of signal recovery from nonadaptive linear measurements can be formulated as a sparse optimization problem which may take different forms, depending on such factors as  recovery environment, signal structure, and the prior information available for the signal \cite{E10,EK12}.
   A fundamental mathematical model for sparsity-based signal recovery can be described as follows.
  Let $A $ be an $m \times n$  sensing  matrix with $m \ll n,$  and let $ y: =Ax +\nu \in \mathbb{R}^m $ be the measurements for the signal $x \in \mathbb{R}^n $, where $\nu\in \mathbb{R}^m $  are the measurement errors. When $ x$ is $k$-sparse or $k$-compressible, the recovery of $x$  can be formulated as the following sparsity-constrained optimization (SCO) problem:
\begin{equation} \label{L0} \min_{z\in \mathbb{R}^n} \{\|Az-y\|_2^2:  ~ \|z\|_0 \leq k\},\end{equation}
 where  $k$ is a given integer number reflecting the number of significant entries of the signal to  recover, and $\|z\|_0$ denotes the number of nonzero entries of the vector $z.$ In this paper, we say that $z$ is $k$-sparse if $ \|z\|_0 \leq k,$ and $z$ is $k$-compressible if  it can be approximated by a $k$-sparse vector. The SCO problem plays a vital role in the development of algorithms for compressed-sensing-based signal recovery (see, e.g., \cite{E10,EK12,FR13,Z18}). Essentially, the model (\ref{L0}) is to find the best $k$-term approximation of the target signal, which best fits the acquired measurements.
Similar models also arise in numerous areas such as statistical learning \cite{M02,BKM16,SBRJ19}, wireless communication \cite{CSDRK17, GDHCWH18}, low-rank matrix recovery \cite{CP11, CZ13, DR16, NNW17, FS19}, linear inverse and optimization problems \cite{DDM04,VW13,W19,ZMRC19}.

While the SCO and related problems can be solved via convex optimization,  nonconvex optimization and orthogonal matching pursuit (OMP) (see, e.g., \cite{E10,EK12,FR13,Z18}),  the thresholding methods with low computational complexity are  particularly suited for solving the SCO model. Thus we focus our attention on the study of thresholding methods in this paper.
The earlier thresholding methods mainly adopt the soft thresholding operator \cite{D95, DJ94, E06, EMSZ07}. The recent ones (e.g., \cite{FR13,HGT06,FR08,BD08,BD10,F11}) mainly use the hard thresholding operator, denoted by $ {\cal H}_k (\cdot),$ which retains the $k$ largest magnitudes of a vector and sets other entries to zeros.
The latest development of thresholding methods, including acceleration, matrix form, dual thresholding and Newton-step-based modifications, can be found in such references as \cite{SBRJ19,FS19,C11,KC14,LYWLM17,KK17,LB19,MZ20}. However, performing hard thresholding on an incompressible iterate, generated by an iterative method, does not ensure the reduction of the objective value of (\ref{L0}) at every iteration.
Motivated by this observation, the so-called optimal $k$-thresholding (OT) method was first introduced in \cite{Z20}, which allows the vector thresholding and objective reduction to be performed simultaneously.  Based on the OT concept,  the relaxed optimal $k$-thresholding (ROT) algorithm  and their variants were also developed in \cite{Z20}. A further analysis of  these algorithms was performed in \cite{ZL21}, and the algorithm using both OT technique and truncated gradient descent direction was studied in \cite{MZMS21}. Compared to   existing hard-thresholding algorithms, the strength of OT-type methods lies in their observed robustness for sparse signal recovery and steadiness in objective reduction in the course of iterations.  However, the OT algorithm needs to solve a quadratic optimization problem at every iteration to obtain the optimal or nearly optimal $k$-thresholding of a vector,  and thus the computation is expensive especially for high-dimensional signal recovery.

This paper pursues a new algorithmic development which is significantly different from the existing ones in \cite{Z20} and \cite{ZL21}. The main idea of the paper is to introduce an auxiliary regularized counterpart for  the so-called binary OT model, which will be defined by (\ref{conrel}) later in this paper.  Based on such a regularized model, we utilize the traditional linearization technique (i.e., the first-order approximation method) to develop a new framework of  algorithms which admits a low computational cost at every iteration, compared to general convex optimization methods. The thresholding technique adopted in this framework is referred to as the \emph{natural thresholding} (NT). Such thresholding is determined by the $k$ smallest entries of the gradient of the so-called regularized objective of OT or ROT model which will be recalled in the next section. The NT algorithm takes the advantage of the `nice structure' of the polytope with $k$-sparse binary extreme points. The main work of the paper includes:
\begin{itemize}

 \item [(i)] Develop a novel class of NT-type methods for signal recovery;

\item[(ii)] Rigorously prove the guaranteed performance of NT-type algorithms under some conditions.
\end{itemize}
We also carry out experiments on random problem instances to show that the NT-type algorithms are fast algorithms, and their performances are very comparable to several mainstream compressed sensing algorithms.

The paper is structured as follows. The NT-type algorithms are proposed in Section \ref{sect2}. The theoretical analysis of the algorithms is performed in Section \ref{sect3}, and a further discussion of   regularization functions and   parameters is given in Section \ref{sect4}. The simulation results are reported in Section \ref{sect5}.

\emph{Notation:} All vectors are column vectors unless otherwise stated. For vector $z$  and matrix $A$,  $ z^T$ and $A^T$  denote their transposes.   We use $\textrm{supp}(z)=\{i: z_i\not=0\}$ to denote the support of the vector $z$, i.e., the index set of nonzero entries of $z. $ We use $L_k (z)$ to denote the index set of the $k$  largest magnitudes  of $z, $ and $ {\cal S}_k(z)$ to denote the index set of the $k$  smallest entries  of $z.$ When $L_k (z)$   ($ {\cal S}_k(z)$) is not uniquely determined, i.e., when there are two magnitudes (entries) being equal, we select the smallest entry indices to avoid the ambiguity in index selection.  Given $ S \subseteq \{1, 2, \dots, n\} , $  we denote by  $ \overline{S}=  \{1, 2, \dots, n\} \backslash S  $  the complement set and by $|S|$ the cardinality of $S.$   Given $ x\in \mathbb{R}^n,$  the vector $ x_S \in \mathbb{R}^n$ is obtained from $ x$ by retaining the entries supported on $S$ and setting other entries to zeros. Throughout the paper,  $\|\cdot\|_2 $ and $\|\cdot\|_1 $ denote the $\ell_2$-norm and $\ell_1$-norm of a vector, respectively.  For vectors $x$ and $ z, $    $x \otimes z$ is the Hadamard product (entry-wise product) of the two vectors.
Also, we summarize the main acronyms of algorithms in the following table.

 \vskip 0.08in
\begin{center}
\begin{tabular}{|c|l|}
 \hline
 Acronyms     &   Algorithms      \\
 \hline
  IHT    &    Iterative hard tresholding   \\
  \hline
  HTP   &     Hard tresholding pursuit  \\
   \hline
 SP &    Subspace pursuit    \\
  \hline
 CoSaMP &    Compressive sampling matching pursuit    \\
 \hline
 OMP    &  Orthogonal matching pursuit                             \\
\hline
  OT  &     Optimal $k$-thresholding    \\
 \hline
  OTP &    Optimal $k$-thresholding pursuit    \\
 \hline
ROTP &    Relaxed optimal $k$-thresholding pursuit    \\
 \hline
 NT  & Natural thresholding   \\
 \hline
  NTP  & Natural thresholding pursuit  \\
 NT$_q$  & Natural thresholding with inner iter. \\
 \hline
 NTP$_q$ & Natural thresholding pursuit with inner iter. \\
 \hline
\end{tabular}
\end{center}

\section{Natural Thresholding Algorithms} \label{sect2}
Let us first recall the iterative hard thresholding (IHT), hard thresholding pursuit (HTP) and the newly developed optimal $k$-thresholding (OT) algorithm. These algorithms provide a basis for the development of our new algorithms called natural thresholding.

\subsection{Basic hard thresholding algorithms: IHT and HTP}

Given $(y, A)$, the gradient of the error metric $\psi(x) = (1/2)\|y-Ax\|_2^2  $ is given by $ \nabla \psi(x) = - A^T (y-Ax). $  At the iterate $ x^{(p)},$ let $ u^{(p)}$ be the vector generated by the classic gradient descent method, i.e., \begin{equation} \label{uupp}  u^{(p)} := x^{(p)} + \lambda_p  A^T (y-Ax^{(p)}), \end{equation}  where $ \lambda_p$ is a steplength. Using the operator ${\cal H}_k$ to generate an iterate satisfying the constraint of the problem (\ref{L0}) leads  to the well-known  IHT algorithm \cite{DDM04,BD08,BD10} and its enhanced version  HTP \cite{F11}. A good feature of IHT and HTP is that they are easier to implement than other algorithms such as convex optimization \cite{CDS98,CT05,CWB08,ZL12}, orthogonal matching pursuit (OMP) \cite{MZ93,PRK93,DMZ94}, compressive sampling matching pursuit (CoSaMP)\cite{NT09}, and subspace pursuit (SP) \cite{DM09}.

\begin{algorithm}

 {\bf IHT algorithm.}  Given the problem data $(A, y, k)$  and initial point $x^{(0)},$ perform the following iteration  until a  stopping criterion is met: At $x^{(p)}$, choose $ \lambda_p > 0$ and set
$$ x^{(p+1)}= {\cal H}_k  (u^{(p)} ), $$
where $ u^{(p)}$ is given by (\ref{uupp}).\\

 {\bf HTP algorithm.}   Given the problem data $(A, y, k)$  and initial point $x^{(0)},$ perform the  following iteration  until a  stopping criterion is met: At $x^{(p)}$, choose $ \lambda_p > 0$ and let $ u^{(p)}$ be given by (\ref{uupp}). Then set
$$ x^+ = {\cal H}_k  (u^{(p)} ), $$
$$ x^{(p+1)} = {\rm arg}\min_{z} \{\|y-Az\|_2^2:  ~ \textrm{supp} (z) \subseteq \textrm{supp} (x^+) \}. $$
\end{algorithm}

Clearly, the $k$ terms of $u^{(p)}$ selected  by ${\cal H}_k$ are not necessarily the best $k$ terms in the sense that they might not be the $k$ terms that best fit the linear measurements. As pointed out in \cite{Z20},  performing the hard thresholding of  $u^{(p)}   $ is independent of the error metric $\psi.$  Thus it may happen that the value of the error metric at the point $ x^+ = {\cal H}_k(u^{(p)})$ is larger than that of the current iterate $x^{(p)},$ i.e., $ \|y-A x^+\|_2 > \|y-Ax^{(p)}\|_2. $
This observation motivates the recent development of optimal $k$-thresholding algorithms in \cite{Z20, ZL21}.

\subsection{Optimal $k$-thresholding algorithms}

As defined in \cite{Z20}, the optimal $k$-thresholding of a vector $u $, denoted by $ Z_k^\# (u ), $   is given by $Z_k^\# (u ) = u \otimes w^*$ which is the Hadamard product of $ u $ and $w^*,$ where $w^*$ is the solution to the problem
 $$  \min_{w}   \left\{  \|y- A (u  \otimes   w) \|^2_2 : ~  \textrm{\textbf{e}}^T w= k ,  ~  w \in \{0,1\}^n \right\},
  $$ where $ \textrm{\textbf{e}}$ is the vector of ones, and the vector $ w $  is   $k$-sparse and binary   with exactly $k$ entries  equal to 1. Given such a binary vector $w,$  the Hadamard product of $ u$ and $ w$ retains $k$ entries of $ u$ only and sets other entries to zeros. Based on the operator $Z_k^\# (\cdot),$ the conceptual optimal $k$-thresholding (OT) method and the optimal $ k$-thresholding pursuit (OTP)  were proposed in \cite{Z20}. The two methods share the same first step except for the second step at which OTP performs an orthogonal projection.

 \begin{algorithm}
 \textbf{OT and OTP algorithms.} Given the problem data  $(A, y, k)  $ and  a starting point $x^{(0)}. $ Perform the following steps until a stopping criterion is satisfied:

 \begin{itemize}
 \item[S1] Let $ u^{(p)} $ be given as (\ref{uupp}) and let $w^*$ be the solution to the problem
\begin{equation} \label{conrel}    \min_{w}   \{  \|y- A (u^{(p)}  \otimes   w) \|^2_2 : ~  \textrm{\textbf{e}}^T w= k , ~    w\in \{0, 1\}^n \}.
\end{equation}

  \item[S2] Generate $x^{(p+1)} $ as follows:
\end{itemize}

\begin{itemize} \item[]  For OT algorithm,  $ x^{(p+1)}=  u^{(p)}  \otimes  w^*.  $
\item[]   For OTP algorithm, let $ \Omega = \textrm{supp} ( u^{(p)}  \otimes  w^*)$ and
 $$ x^{(p+1)}= \textrm{arg}\min_{z} \{ \|y-A z\|_2^2:  ~ \textrm{supp} (z)\subseteq  \Omega   \}.  $$
\end{itemize}
\end{algorithm}
We would rather treat OT and OTP as conceptual algorithms due to the fact that directly solving the binary problem (\ref{conrel}) might be  expensive when the dimension of the problem is high. A more practical way is to solve a convex relaxation of (\ref{conrel}), which can be solved by existing interior-point methods.  This consideration leads to the relaxed optimal $k$-thresholding (ROT) and relaxed optimal $k$-thresholding pursuit (ROTP) introduced in \cite{Z20, ZL21}.

\begin{algorithm}
 \textbf{ROT and ROTP algorithms.} Given the problem data  $(A, y, k)  $ and  a starting point $x^{(0)}, $ perform the  steps below until a   stopping criterion is satisfied:

 \begin{itemize}
 \item[S1] Let $u^{(p)} $ be given as (\ref{uupp}) and solve
\begin{equation} \label{relax}    \min_{w}   \{  \|y- A (u^{(p)}  \otimes   w) \|^2_2 : ~  \textrm{\textbf{e}}^T w= k , ~  0\leq w\leq \textrm{\textbf{e}} \}.
\end{equation} Let $w^{(p)}$ be the solution to this problem.

  \item[S2] Set  $ \widehat{x} =  {\cal H}_k (u^{(p)}  \otimes  w^{(p)}) $   and generate $x^{(p+1)} $ as follows:
\end{itemize}

\begin{itemize} \item[]  For ROT algorithm,  $ x^{(p+1)}=   \widehat{x}.  $

\item[]   For ROTP algorithm,  let
 $$ x^{(p+1)}= \textrm{arg}\min_{z} \{ \|y-A z\|_2^2:  ~ \textrm{supp} (z)\subseteq  \textrm{supp}(\widehat{x}) \}.  $$

\end{itemize}
\end{algorithm}

While the optimization problem (\ref{relax}) can be solved efficiently by existing convex optimization solvers, the computational cost remains high for large-scaled signal recovery problems, compared with the algorithms which involve only orthogonal projection, hard thresholding or linear programming solving. This stimulates the natural thresholding (NT) method described in the next subsection.

\subsection{Natural thresholding algorithms: A new framework}

The model (\ref{relax}) is a convex relaxation of (\ref{conrel}).  However, such a relaxation might be too loose, and thus its solution is usually not binary.  To overcome such a drawback of  (\ref{relax}), we may introduce a certain regularization term into the objective of (\ref{relax}) so that the regularized model can approximate the model (\ref{conrel}) appropriately and the chance for the solution of regularized model being binary is enhanced.  We now describe such a regularized model which is essential for the development of new algorithms in this paper.
 In fact, given vector $ u^{(p)},$ consider the model of the form:
  \begin{eqnarray}   \label {HT-QP-Relax-2}  & \min  &    \|y- A (u^{(p)}  \otimes   w) \|^2_2 + \alpha \phi(w)  \\
& {\rm s.t.} &  \textrm{\textbf{e}}^T w= k , ~  0\leq w\leq \textrm{\textbf{e}},   \nonumber
  \end{eqnarray}
  where $\alpha>0$ is a parameter and   $\phi(w)$ is referred to as  a \emph{binary regularization} function  which is formally defined as follows.

  \begin{Def} \label{Def01} $\phi  $ is said to be a binary regularization if it satisfies the following two properties: i) $\phi$ is a
  positive and continuously differentiable function over an open neighborhood of the region $ D:= \{ w\in \mathbb{R}^n: 0\leq w\leq \textrm{\textbf{e}}\};$  ii) $\phi(w)$ attains its minimum value over $D$ at any binary vector $ w\in \{0, 1\}^n$ and only at such binary vectors.
\end{Def}

\begin{Exa} Note that given any positive number $ 0 < \sigma <1$, the function $ \tau(t): = (t+\sigma) (1+\sigma -t)$  is positive and continuously differentiable over the interval $ -\sigma <t <1+\sigma$ which is an open neighborhood of the interval [0,1].  $\tau (t) $  attains its minimum value $ \tau_{\min} := \sigma (1+\sigma)$ at $t=0 $ or $1, $ and $ \tau (t) > \tau_{\min}  $ for any $ 0< t< 1.$  For simplicity,  we set $\sigma =1/2$ and only consider the specific function $ \tau (t) = (t+\frac{1}{2} ) (\frac{3}{2}  -t)  .$
By using this univariate function, we may construct a binary regularization function in $\mathbb{R}^n$ as follows:
  \begin{equation} \label{phi}  \phi (w)  = \sum_{i=1}^{n} \tau (w_i)= \sum_{i=1}^{n} (w_i+\frac{1}{2}) (\frac{3}{2} -w_i).
  \end{equation}
  It is evident that the function (\ref{phi}) satisfies the properties specified in Definition \ref{Def01}.  More examples of  binary regularization functions will be given later in this section.
\end{Exa}

 Given a vector $ u^{(p)},$  we use $f$ and $ g_\alpha $ to denote the  functions
$$f(w) = \|y-A(u^{(p)}\otimes w)\|_2^2 ,$$
\begin{equation} \label{ggaa}  g_\alpha (w) =  f(w) + \alpha \phi (w).  \end{equation}
  The function $ g_\alpha (w) $ is the objective of the problem (\ref {HT-QP-Relax-2}). The parameter $\alpha$ is referred to as the \emph{regularization parameter}.
The solution of (\ref{HT-QP-Relax-2}) depends on the parameter $ \alpha.$ Thus for any given $ \alpha>0,$ we use $ w(\alpha) $ to denote the solution of (\ref{HT-QP-Relax-2}).   $\phi(w)$ can be seen as the penalty for the variable $w$ violating the 0-1 restriction. From the classic optimization theory for penalty functions (see, e.g., \cite{FM68}),  it is not difficult to see that as $ \alpha \to \infty,$ the value of $ \phi(w(\alpha))$ will decrease to its minimum in the feasible set of (\ref{HT-QP-Relax-2}). Such minimum attains only at 0-1 vectors according to the definition of binary regularization. This implies that
 $ w(\alpha)$  tends to a 0-1 point in $D$ as $ \alpha \to \infty,$ and any accumulation point of $ w(\alpha)$ must be a solution to the  problem (\ref{conrel}). We summarize this fact in the following lemma whose proof is given in {\sc Appendix A} for completeness.

 \begin{Lem} \label{Lemma-penality} Let $ \phi(w) $ in (\ref{HT-QP-Relax-2})  be  a binary regularization and $ w(\alpha) $ be a solution of (\ref{HT-QP-Relax-2}). Consider any strictly increasing and positive sequence $ \{\alpha_{\ell}\} $ satisfying $\alpha_{\ell} \to \infty.$   Then the sequence $ \{\phi(w(\alpha_{\ell}))\}$ is nonincreasing and converges to the minimum value of $ \phi $ over $D,$ and the sequence $\{f(w(\alpha_{\ell}))\}$ is nondecreasing and converges to the optimal value of (3). Moreover, any accumulation point of $ \{w(\alpha_{\ell})\}$ is the optimal solution of (3).
 \end{Lem}

 The above lemma shows that the model (\ref{HT-QP-Relax-2}) can approximate the model (\ref{conrel}) to any expected level of accuracy provided that the parameter $ \alpha $ is taken large enough. From this perspective, the model (\ref{HT-QP-Relax-2}) as an approximation of (\ref{conrel}) is better than the convex relaxation (\ref{relax}).  By the definition, a binary regularization is nonconvex. As shown by (\ref{phi}), it is very easy to construct a binary regularization which is concave. As a result, the objective function $g_\alpha $ of (\ref{HT-QP-Relax-2}) can be naturally made to be concave by properly choosing the parameter $ \alpha.$  Thus for the convenience of discussion, we assume that $g_\alpha $ is concave throughout the remainder of this section and Section \ref{sect3}. The concavity of the function $ g_\alpha(w)$ can be guaranteed if   $ \phi $  and $ \alpha$ are chosen properly.   Note that if $ \phi$ is twice continuously differentiable, then the Hessian of $g_\alpha  $ is given as
 $$   \nabla^2  g _\alpha (w) =  \nabla^2 f(w)+ \alpha  \nabla^2  \phi(w), $$
 from which it can be seen that when  $  \nabla^2  \phi(w)$ is negative definite,  then $\nabla^2  g _\alpha (w)$ will be negative semi-definite (and thus $ g_\alpha$ is concave) provided that $ \alpha$ is large enough.

 Take (\ref{phi}) as an example. It is easy to verify that the Hessian of (\ref{phi}) is   $\nabla^2  \phi (w)= -2I,$ where $I$ is the identity matrix. Thus when (\ref{phi}) is used, the Hessian of $g_\alpha(w)$ is given as
 $$ \nabla^2  g _\alpha (w)    = 2 (U^{(p)} A^T A U^{(p)} - \alpha I), $$
  where $U^{(p)}=  \textrm{diag} (u^{(p)}).$
 Note that  $ \nabla^2  g_\alpha (w)$ is negative semi-definite over the whole space provided that $ \alpha I\succeq U^{(p)} A^T A U^{(p)} $ which can be guaranteed when $ \alpha$ is not smaller than the largest eigenvalue of the matrix $U^{(p)} A^T A U^{(p)},$ i.e.,  $\alpha \geq \alpha^*:= \lambda_{\max} (U^{(p)} A^T A U^{(p)}). $
 Therefore, the function $g_\alpha (w)$ is always concave over $\mathbb{R}^n  $ provided that $ \alpha \geq \alpha^*.$

  Note that (\ref{conrel}) is to find the optimal $k$-thresholding of $ u^{(p)}.$   The above discussion indicates that performing the optimal $k$-thresholding of a vector can be approximately achieved by solving the problem (\ref{HT-QP-Relax-2}) which from the above discussion can be made to be a concave minimization problem which remains  NP-hard in general  \cite{B95,BT18}. The fundamental idea of the traditional MM method for solving a nonconvex optimization problem is to  replace the `difficult' objective function of the problem with a certain relatively easy surrogate convex function that dominates the original objective function (see, e.g., \cite{MK08, L16}). For a concave objective, the simplest surrogate   function is its first-order approximation at any given point. Thus the first-order approximation of the concave function can be used to develop an algorithm with significantly low computational cost.

Specifically, suppose that $w^-$  is the current iterate.    At $ w^-,$ by the concavity of $g_\alpha (w) $ over an open neighborhood of  $D,$ denoted by $O,$ one has
\begin{equation} \label{IINNEE} g_\alpha (w)  \leq g_\alpha (w^-) + \left[\nabla g_\alpha (w^-)\right]^T (w-w^-)
  \end{equation}
  for any $ w\in O $, where $\nabla g_\alpha (w^-)$ denotes the gradient of $ g_\alpha $ at $ w^-.$  Specifically,  if $\phi $ is given by (\ref{phi}), the gradient of $ g_\alpha  $ at $ w^-$ is given as
    $$   \nabla  g_\alpha (w^-)   =   -2 U^{(p)} A^T (y-AU^{(p)} w^-)+\alpha (\textbf{\textrm{e}} - 2 w^-).
$$
   The relation (\ref{IINNEE})  shows that the first-order approximation of $ g_\alpha(w)$ dominates   $ g_\alpha(w)$  on the whole open set $O.$  By the idea of MM methods, the problem (\ref{HT-QP-Relax-2}), i.e.,  $ \min \{g_\alpha (w): w\in P\},$ where
     $$ P=\{ w\in \mathbb{R}^n:  \textrm{\textbf{e}}^T w =k, ~ 0\leq w\leq \textrm{\textbf{e}}\}, $$
     can be approximated by its surrogate counterpart which, in this case, is the linear programming (LP) problem
  $$ \min\left\{g_\alpha (w^-) + \left[\nabla g_\alpha (w^-)\right]^T (w-w^-): ~ w\in P \right\}. $$
  Removing constant terms in the objective above, the above  LP problem is reduced to
\begin{equation} \label{LP} \min \left\{\left[\nabla g_\alpha (w^-)\right]^T w:   ~ w\in P  \right \}. \end{equation}
Let $ w^+$ be the solution to this problem. We may replace $ w^-$ with $ w^+$ and repeat the above process until a certain stopping criterion is met. This is the basic idea for the development of a new algorithm.

We now point out that there is no need to solve the LP problem (\ref{LP}) by an algorithm since an optimal solution of (\ref{LP}) can be obtained explicitly due to the nature of the polytope $P.$  In fact, according to the classic LP theory,  the problem (\ref{LP}) must have an optimal solution, denoted by $w^+,$ which is an extreme point of the polytope $P. $ As shown in \cite{Z20},  any extreme point of $P$ is $k$-sparse and binary. Hence the optimal value $ \left[\nabla g_\alpha (w^-)\right]^T w^+$ is actually the summation of $k$ entries of the vector $\nabla g_\alpha (w^-).$  Clearly, the smallest value of the function $ \left[\nabla g_\alpha (w^-)\right]^T w$ over the polytope $P$ is the sum of the smallest $k$ entries of $\nabla g_\alpha (w^-).$  Thus the optimal solution $w^+$ of the LP problem (\ref{LP}) can be explicitly given as follows. It is determined by the support of the $k$ smallest entries of  $\nabla g_\alpha (w^-)$, i.e.,
$     w^+_i=   1 $   if  $  i \in   {\cal S}_k (\nabla g_\alpha (w^-)) $, otherwise, $     w^+_i=   0 $
 for $ i =1, \dots, n, $
where $ {\cal S}_k (\cdot) $ is the index set of the $k$ smallest entries of a vector. We now describe the algorithms which are referred to as the natural thresholding (NT) and the natural thresholding pursuit (NTP)  in this paper.

\begin{algorithm}
\caption*{{\bf NT and NTP algorithms.}} Input $(A, y, k)  $ and  an initial point $x^{(0)}  $.  Repeat the following steps until a certain stopping criterion is satisfied:
 \begin{itemize}
 \item[S1]  At  $x^{(p)}$, choose $ \lambda_p >0$ and let $u^{(p)}   $  be given by (\ref{uupp}). Let $w^- $ be the $k$-sparse vector given as
  \begin{equation}\label{www-}  w^-_i= \left\{ \begin{array}{ll} 1 & \textrm{ if }  i \in   L_k (u^{(p)})\\
0 & \textrm{ otherwise}
\end{array},  ~ i =1, \dots, n. \right.   \end{equation}
Compute the gradient $ \nabla g_\alpha (w^-),$ where $ g_\alpha (\cdot)$ is defined by (\ref{ggaa}).
Then set $ w^+ $ as
 \begin{equation}  \label {www+} w^+_i= \left\{ \begin{array}{ll} 1 & \textrm{ if }  i \in  {\cal S}_k (\nabla g_\alpha (w^-))\\
0 & \textrm{ otherwise}
\end{array},  ~ i =1, \dots, n. \right.   \end{equation}

\item[S2] Generate $x^{(p+1)} $ according to  the following procedure:
 \end{itemize}

 \begin{itemize}
 \item[] For NT algorithm,  $ x^{(p+1)}=   u^{(p)}  \otimes  w^+.  $
 \item[]  For NTP algorithm,  set $ S^{(p+1)}= \textrm{supp} ( u^{(p)}  \otimes  w^+  ) $
 and
 $$x^{(p+1)} = \textrm{arg} \min_{z}\{ \|y-A z\|_2^2:  ~ \textrm{supp} (z)\subseteq S^{(p+1)} \}. $$

 \end{itemize}
\end{algorithm}

The only difference between the two algorithms is that NT does not perform an orthogonal projection, but NTP does. At the beginning of every iteration of  NT and NTP, the initial point $ w^-$ can be any given $k$-sparse and binary vector. In particular, we may use the binary vector (\ref{www-}) as an initial point determined by $L_k(u^{(p)})$.

\begin{Rem} \label{Lem22}   Given a $k$-sparse and binary vector $ w^-$, let $ w^+$ be given as (\ref{www+}) which is an explicit solution of the LP problem (\ref{LP}). By the optimality of $ w^+,$ one always has
\begin{equation}\label{DEC}  \left[\nabla g_\alpha (w^-)\right]^T w^+ \leq  \left[\nabla g_\alpha (w^-)\right]^T w^-,  \end{equation}
\end{Rem}
which together with the concavity of $ g_\alpha $ implies that $ f(w^+) \leq f(w^-).$ (This fact will be used in the proof of Lemma \ref{Lem-a} later.) In fact, the concavity of $ g_\alpha(w)$ and (\ref{DEC}) implies that
 $$g_\alpha (w^+) \leq g_\alpha(w^-) + \nabla g_\alpha (w^-) ^T (w^+-w^-) \leq g_\alpha(w^-), $$
 i.e., $$  f(w^+) + \alpha \phi(w^+) \leq f(w^-)+ \alpha \phi(w^-). $$ Since $w^+$ and $w^-$ are 0-1 vectors at which $\phi$ achieves the minimum value $\phi(w^+)=\phi(w^-) $ over $D.$ Thus the above inequality reduces to  $ f(w^+) \leq f(w^-).$
 This means the iterate $w^+$ generated by NT or NTP is never worse than the one generated by IHT or HTP from the perspective of the error metric $f(w).$
 If  $w^-$  is not optimal  to (\ref{LP}), then $ \left[\nabla g_\alpha (w^-)\right]^T w^+ < \left[\nabla g_\alpha (w^-)\right]^T w^- ,$ and hence  $ f(w^+)< f(w^-).$ Thus in this case the algorithm finds the $k$ terms of $u^{(p)}$ which is better than the $k$ terms selected by the hard thresholding.

The NT and NTP algorithms perform linearization of $g_\alpha$ only once in every iteration. We may perform  linearization more than once within their step S1,  i.e.,    replacing the role of $ w^-$ with $w^+$ and repeating the step S1 of NT more than once so that a point better than $ w^+ $ might be found. Such iterations within  S1 are referred to as \emph{inner iterations}.  Repeating the inner iterations up to $ q\geq 1$ times yields the algorithms NT$_q$ and NTP$_q$.

\begin{algorithm}
\caption*{{\bf NT$_q$  and NTP$_q$ algorithms.}} Input $(A, y, k)$,  initial point $x^{(0)} ,$ and integer number $ q\geq 1.$  Repeat the steps below until a stopping criterion is met:

 \begin{itemize}
 \item[S1]   At  $x^{(p)}$, choose $ \lambda_p>0 $ and let $u^{(p)} $ be given by (\ref{uupp}).
 Let $w^- $ be the $k$-sparse vector given by (\ref{www-}).
Then perform the following inner iterations:

 \begin{itemize} \item[]  \textbf{For} $J=1:q$

   \begin{itemize}  \item[]  Compute $ \nabla  g_\alpha(w^-)$, where $ g_\alpha (\cdot)$ is defined by (\ref{ggaa}). Then set $ w^+ $ as (\ref{www+}).

  \item [] \begin{itemize} \item[] \textbf{If}  $ \left[\nabla g_\alpha (w^-)\right]^T w^+ = \left[\nabla g_\alpha (w^-)\right]^T w^-  $ \textbf{then}

~~ terminate the inner iterations, and go to S2

\item[] \textbf{else}

 ~~ $ w^-:= w^+ $

 \item[] \textbf{endif}
 \end{itemize}
  \end{itemize}
  \textbf{End}
 \end{itemize}

\item[S2] Generate $x^{(p+1)} $ according to  the following procedure:
 \end{itemize}
 \begin{itemize}
 \item[]  For NT$_q$ algorithm, set $ x^{(p+1)}=   u^{(p)}  \otimes  w^+.  $
 \item[]  For NTP$_q$ algorithm, let $ S^{(p+1)} = \textrm{supp} (   u^{(p)}  \otimes  w^+ ) $
 and
 $$ x^{(p+1)}= \textrm{arg}\min_{z}\{ \|y-A z\|_2^2:  ~ \textrm{supp} (z)\subseteq S^{(p+1)} \}. $$
 \end{itemize}

\end{algorithm}

\begin{Rem} NT, NTP, NT$_q$ and  NTP$_q$  are referred to as the NT-type algorithms in this paper. When $q=1,$ these algorithms become NT and NTP, respectively.  If $q$ is  a large number, according to linearization theory for concave minimization (e.g., \cite{B95,M96}), the inner iterations of NT$_q$ and NTP$_q$  will terminate at a stationary point of the concave minimization problem in which case $ \left[\nabla g_\alpha (w^-)\right]^T w^+ = \left[\nabla g_\alpha (w^-)\right]^T w^- .$  If we expect such a point, we may set $ q=\infty$ and use $ \left[\nabla g_\alpha (w^-)\right]^T w^+ = \left[\nabla g_\alpha (w^-)\right]^T  w^- $ as the termination criterion for inner iterations.
 The corresponding algorithms can be referred to as NT$_\infty$ and NTP$_\infty$ which are omitted here.
\end{Rem}

 \begin{Rem} As pointed out in \cite{ZL21}, performing one iteration of ROTP requires about $ O( m^3  + mn + n^{3.5} L)$ flops, where $L$ is the size of the problem data encoding in binary. The complexity of NT-type algorithms is much lower than that of ROTP. In fact, at every iteration, the NT-type algorithms need to compute the vector $u^{(p)}$ and $ \nabla g_\alpha (w^-)$ which require about $O(mn)$ operations, and to perform $L_k $ and $ {\cal S}_k$ on a vector which require about $ O(n \log k) $ flops by a sorting approach, where $ k < m \ll n .$   The NTP and NTP$_q$ algorithms need to perform the orthogonal projection $\min\{\|y-Az\|_2^2: \textrm{supp} (z) \subset \Omega \}, $ where $ |\Omega | = l \leq m, $ which requires about $ m^3 $ flops (see \cite{ZL21} for details).  Thus the complexities of NT and NTP are about $ O(mn)$ and  $ O( m^3 + mn) $ flops in every iteration, respectively. If the inner iterations are executed $q$ times, then the NT$_q$ and NTP$_q$ require about
 $ O(qmn)$ and  $ O( m^3 + qmn) $ flops.
\end{Rem}

\subsection{Concave regularization function}
Before we analyze the guaranteed performance of NT-type algorithms, it is necessary to construct more examples of concave binary regularization functions.
We still construct such functions based on the kernel univariate one $\tau (t) = (t+1/2)(3/2-t) $ which is   concave and positive  over an open neighborhood of the interval $ [0 ,  1].$   Assume $ g $ is another concave and strictly increasing univariate function over the interval $(0, \infty). $    Then the composite function
 $ h(t)= g(\tau(t))$ must be a concave function over an open neighborhood of the interval $ [0, 1]$  due to the  fact
 $$ h''(t) = g'' (\tau (t)) ( \tau'(t))^2 + g'(\tau(t)) \tau''(t) \leq  0,  $$
 where the last inequality follows from the fact $ g'' (\tau (t)) \leq 0, \tau''(t) \leq  0$ and $ g'(\tau(t))\geq 0.$
 We may use such a  function  to construct a binary regularization as follows:
 \begin{equation} \label{gghh}  \phi(w)= \sum_{i=1}^n h(w_i) = \sum_{i=1}^n g(\tau(w_i)). \end{equation}
 The function constructed like this can be used  in NT-type algorithms.
We now give two such specific functions.
\begin{itemize}

\item Let $  g (t):  =\log (1+t) $ which is concave and strictly increasing over the interval $ (0, \infty). $   Then
\begin{equation}  \label{RG02}   \phi(w) = \sum_{i=1}^n \log \left(1+ (w_i+1/2)(3/2-w_i)\right)   \end{equation}
is a concave binary regularization   over an open neighborhood of the region $ D. $

\item  Let $  g (t):  = \frac{t}{1+t}  $ which is concave and strictly increasing  over the interval $ (0, \infty).$  Then
\begin{equation}  \label{RG03}  \phi(w) = \sum_{i=1}^n  \frac{(w_i+1/2)(3/2-w_i)}{ 1+ (w_i+1/2)(3/2-w_i)}  \end{equation}
 is a concave binary  regularization  over an open neighborhood of the region $D . $

 \end{itemize}

 Similar to $f (w) $ which is related to $u^{(p)}$,  the regularization $ \phi$ can be also  made  related to $ u^{(p)}.$ Such a regularization will be discussed in more detail in Section \ref{sect4}.

\section{Guaranteed performance and stability} \label{sect3} In this section, we perform a theoretical analysis for  the NT-type algorithms described in previous section. This analysis is made under the assumptions of concavity of $ g_\alpha (w)$ and restricted isometry property (RIP), which is initially introduced by Cand\`es and Tao \cite{CT05}.
An $ m \times n$  matrix $A,$ where $m \ll n,$ is said to satisfy the RIP of order $K$  if $ \delta_K < 1 ,$ where $\delta_K $ is  the smallest nonnegative number $ \delta$  such that
$$ (1-\delta) \|z\|_2^2  \leq \|Az\|^2_2 \leq (1+ \delta) \|z\|_2^2  $$  for any $K$-sparse vector $z\in \mathbb{R}^n.$ The number $\delta_K $ is called the restricted isometry constant (RIC) of $A.$
The following property of $ \delta_K$ has been widely used in the analysis of compressed sensing algorithms.

\begin{Lem} \label{333111}  \cite{CT05, FR13} Let $u\in \mathbb{R}^n$ and $\Lambda \subseteq \{1, 2, \dots, n\},$ if $|\Lambda \cup \textrm{supp}(u)| \leq t,$ then $\|\left[(I-A^TA )u\right]_\Lambda \|_2 \leq \delta_t \|u\|_2. $
\end{Lem}

It is worth stressing that  RIP is not the unique assumption under which the compressed sensing algorithms can be analyzed. Other assumptions such as the mutual coherence \cite{E10, FR13}, null space property \cite{FR13}, and  range space property of $ A^T$ \cite{Z18, Z13} can be also used to analyze the theoretical performance of a compressed sensing algorithm.
We now prove that the iterates generated by the NT-type algorithms  with a concave binary regularization can guarantee to recover the sparse signal under the RIP assumption.  We begin with the following lemma.

\begin{Lem} \label{Lem-a}    Let $g_\alpha (w)$ be a concave function, and let $w^-$ be the initial point of step S1 in  NT$_q$ and NTP$_q$  algorithms where $ q\geq 1, $ and let $ w^+$ be the final output of step S1.
Then
\begin{equation} \label{DDEE} \| y- A (u^{(p)} \otimes w^+)\|_2 \leq \| y- A (u^{(p)} \otimes w^-)\|_2 . \end{equation}
\end{Lem}

\emph{Proof.} The inner iterations within step S1 of NT$_q$ and  NTP$_q$ start from the initial point $w^{(0)}:= w^-.$ Then the first inner iteration performs the update (\ref{www+}) to yield the next inner iterate $w^{(1)} .$ As pointed out in Remark \ref{Lem22}, since $ g_\alpha(w)$ is concave and  $ (w^{(0)}, w^{(1)})$  are 0-1 vectors, one must have that $ f(w^{(1)}) \leq f(w^{(0)}). $  Then the inner iteration starts from $ w^{(1)}$ to generate the second iterate $w^{(2)}. $  Again, by the same argument in Remark \ref{Lem22}, one has $ f(w^{(2)}) \leq f(w^{(1)}). $ Repeating this inner process $q$ times, the final output $w^{(q)} $ of step S1 satisfies that
$$   f(w^{(q)}) \leq  f(w^{(q-1)}) \leq  \cdots \leq  f(w^{(1)}) \leq  f(w^{(0)})  . $$
Denote by $w^+ (=w^{(q)})$ the final output of S, the above inequality  immediately implies the desired relation (\ref{DDEE}).   \hfill $ \Box $

The next lemma is very useful in our analysis.

\begin{Lem} \label{Lem4433} \cite{ZL21}  For any vector $z \in \mathbb{R}^n $ and  any $k$-sparse vector $h\in \mathbb{R}^n,  $    one has
$$ \| ( h -   {\cal H}_k(z))_{S\backslash S^*} \|_2 \leq   \|(h-z )_{ S \backslash S^*}\|_2+  \|(  h -   z)_{S^*\backslash S} \|_2 ,  $$
where  $S=  \textrm{supp}  (h)  $ and  $   S^* =  \textrm{supp}  ( {\cal H}_k(z) ). $
\end{Lem}

We now prove the guaranteed performance of the NT-type algorithms under RIP and concavity of $ g_\alpha(w).$
It is sufficient to show the main result in noisy settings. The result for noiseless cases
can be obtained immediately as   a special case  of our results.
The practical signal is usually not exactly $k$-sparse. Thus we are interested in recovering
the key information of the signal, i.e., the largest $k$ magnitudes of the signal.    For simplicity, we only analyze the algorithms with
steplength $ \lambda_p \equiv 1. $ The performance of the proposed algorithms with $ \lambda_p \not=1$ may
be established by a more complicated analysis which requires significant further investigation. In fact, for the case $ \lambda_p \not=1,$ the term $ I-\lambda_p A^TA $ would appear in (\ref{ER02}) in the proof of  Theorem \ref{Thm01} below. See {\sc Appendix B} for details. Thus, to bound the term  in (\ref{ER02}), the classic Lemma \ref{333111}  cannot be used, and some new technical results  involving $ \lambda_p \not=1 $ must be established first.
 We leave such a possible generalization as a future research work.

\begin{Thm} \label{Thm01}  Let $ y: = Ax +\nu $ be the acquired measurements for the signal $x,$ where $\nu$ are measurement errors. Denote by $ S= L_k(x). $ Suppose that the RIC of $ A$ satisfies
    \begin{equation} \label{4A}  \delta_{3k} < \gamma^* \approx 0.4712 , \end{equation}   where $ \gamma^*$ is the unique real root   of the univariate equation $  \omega \gamma ^3 + \omega  \gamma ^2 + \gamma =1  $ in the interval $[0,1],$ where $ \omega: = (\sqrt{5}+1)/2.$
     Suppose that the sequence  $ \{ x^{(p)} : p\geq 1\} $ is generated by the algorithm NT or NT$_q$ with $ \lambda_p \equiv 1  $ and $ g_\alpha (\cdot) $ being concave. Then the sequence $ \{ x^{(p)}  \} $  satisfies the following relation:
    \begin{equation} \label{EERRBB}  \| x^{(p+1)}-x_S \|_2 \leq \eta \|x^{(p)}-x_S \|_2 + C_1 \|A^T  \nu' \|_2+ C_2\|\nu'\|_2,  \end{equation}
    where
    \begin{equation} \label{C1C2} C_1= \omega \sqrt{\frac{ 1+\delta_{2k}}  {1-\delta_{2k}}} , ~  C_2= \frac{2}{\sqrt{1-\delta_{2k}}}, ~ \nu' = A x_{\overline{S}}+ \nu,
    \end{equation}
    and    $  \eta    $  is a constant given by
   \begin{equation}  \label{RE}  \eta= \omega\delta_{3k} \sqrt{ \frac{1+\delta_{2k}}{1-\delta_{2k}}   }  \end{equation}
which is smaller than 1 under the condition (\ref{4A}).
    \end{Thm}

The proof of Theorem \ref{Thm01} is given in {\sc Appendix B}. It should be pointed out that to ensure the concavity of $ g_\alpha (\cdot)$, one may use concave binary regularization and choose $ \alpha $ to be large enough (see Section \ref{sect4} for more discussions).   The only difference between NT$_q$  and  NTP$_q$ is that the latter uses the orthogonal projection to generate  $ x^{(p+1)}$ from  $ x^{(p)}. $ The following property of orthogonal projection is widely used in the analysis of compressed sensing algorithms.

 \begin{Lem} \cite{Z20, ZL21} \label{Orth} Given the pair $(y,A)$ with $ \nu'=  y- A x^*$ where $ x^*$ is a $k$-sparse vector.
Let  $u \in \mathbb{R}^n$ be an arbitrary $k$-sparse vector, and
let $z^*$ be  the   solution to the orthogonal projection
$$ z^{*}=\arg \min _{z}\left\{\|y-A z\|_{2}^{2}: ~ \operatorname{supp}(z) \subseteq \operatorname{supp}(u)\right\}. $$
Then $ z^*$ satisfies that
$$ \left\|z^{*}-x^*\right\|_{2} \leq \frac{1}{\sqrt{1-\delta_{2 k}^{2}}}\|x^*-u\|_{2}+\frac{\sqrt{1+\delta_{k}}}{1-\delta_{2 k}}\|\nu'\|_{2} . $$
\end{Lem}

Combining Theorem \ref{Thm01} and Lemma \ref{Orth} leads to the main result for NTP and   NTP$_q. $

 \begin{Thm} \label{Thm02}  Let $ y: = Ax +\nu $ be the  measurements of the signal $x$ with measurement errors $\nu.$  If the RIC   of $A$ satisfies
    \begin{equation} \label{4D}     \delta_{3k} < \frac{2} { \sqrt{5}+3 }  , \end{equation}
    then the sequence $ \{x^{(p)}: p\geq 1\},$ generated by the algorithm NTP or NTP$_q$ with  $ \lambda_p\equiv 1$ and $ g_\alpha (\cdot)$ being concave, satisfies the following relation:
 \begin{equation} \label{4E}
     \| x^{(p+1)}-x_S \|_2 \leq   \rho  \|x^{(p)}-x_S \|_2 + \widetilde{C}_1 \|A^T \nu' \|_2+\widetilde{C}_2\|\nu'\|_2,
     \end{equation}
    where $ \nu'=A x_{\overline{S}} +\nu,$    $    \rho   $  is a  constant given by
   $    \rho   =  \omega
     \delta_{3k} /(1-\delta_{2k})   $
which is smaller than 1 under (\ref{4D}), and $ \widetilde{C}_1,$ $\widetilde{C}_2$ are the constants given by
\begin{equation}\label{CC33} \widetilde{C}_1= \frac{C_1}{\sqrt{1- \delta_{2 k}^{2}}}, ~
  \widetilde{C}_2= \frac{C_2}{\sqrt{1- \delta_{2 k}^{2}}}  + \frac{\sqrt{1+\delta_k}}{1-\delta_{2k}},
 \end{equation}
   in which $ C_1, C_2$ are the constants given in Theorem \ref{Thm01}.

    \end{Thm}

\emph{Proof.} By the structure of NTP and NTP$_q,$  $ w^+$ is the output of S1, and $ \hat{x} = u^{(p)} \otimes w^+$ is $ k$-sparse. The iterate $ x^{(p+1)} $ is the solution to the orthogonal projection
$$   \min _{z}\left\{\|y-A z\|_{2}^{2}: ~ \operatorname{supp}(z) \subseteq \operatorname{supp}(\hat{x})\right\}. $$
Let $ S= L_k(x).$ By treating $ u=\hat{x}$ and $ x^* = x_S$ as well as $ \nu'= y-Ax_S=Ax_{\overline{S}}+
\nu $,
it follows from Lemma \ref{Orth} that
$$    \|x^{(p+1)}-x_S  \|_{2} \leq \frac{\|x_S -\hat{x}\|_{2}}{\sqrt{1-\delta_{2 k}^{2}}}+\frac{\sqrt{1+\delta_{k}}}{1-\delta_{2 k}}\|
\nu'\|_{2} . $$
The intermediate point $ \hat{x},$ generated at the $p$th iteration of  NTP (NTP$_q$)  are exactly the iterate generated at the $p$th iteration of  NT (NT$_q$).
 By Theorem \ref{Thm01}, we have
 $$ \| x_S-\hat{x} \|_2    \leq  \eta  \|x_S-x^{(p)}\|_2  + C_1\|A^T\nu'\|_2+ C_2\|\nu'\|_2, $$
 where $ \eta $, $C_1, C_2$ are given in Theorem \ref{Thm01}.
 Combining the above two relations, we immediate obtain that
 \begin{align*}      \|x^{(p+1)}     -x_S  \|_{2}    \leq  \rho   \|x_S-x^{(p)}\|_2  + \widetilde{C}_1 \|A^T \nu'\|_2+  \widetilde{C}_2 \|\nu'\|_2  ,
 \end{align*}
 where $$ \rho    =   \frac{\eta}{\sqrt{1-\delta_{2 k}^{2}}}     = \frac{\omega \delta_{3k} }{\sqrt{1-\delta_{2 k}^{2}}}  \sqrt{ \frac{ 1+\delta_{2k}}{1-\delta_{2k}} }    = \frac{\omega \delta_{3k} }{1-\delta_{2k}}  , $$
 and $$
 \widetilde{C}_1 = \frac{C_1}{\sqrt{1- \delta_{2 k}^{2}}}, ~  \widetilde{C}_2= \frac{C_2 }{\sqrt{1- \delta_{2 k}^{2}}}  + \frac{\sqrt{1+\delta_k}}{1-\delta_{2k}}.  $$
Since $ \delta_{2k} \leq \delta_{3k},$ one has $   \rho \leq \frac{\omega \delta_{3k} }{1-\delta_{3k}}.$ Thus $ \rho <1 $ is guaranteed if $ \frac{\omega \delta_{3k} }{1-\delta_{3k}} <1,$ which is ensured under the condition
$  \delta_{3k} < \frac{1} { 1+ \omega}  =\frac{2}{3+\sqrt{5}}. $
Thus the  bound (\ref{4E}) is valid.  \hfill $ \Box $

When the measurements are accurate and the signal is $k$-sparse, we immediately have following corollary.

\begin{Cor} \label{Cor37} Suppose that the signal $ x$ is $k$-sparse and $ y: =Ax. $ Then the following two statements are valid:

(i)  If the RIC of $A$ satisfies (\ref{4A}), then the sequence $ \{x^{(p)}: p \geq 1\}, $ generated by  NT or NT$_q$ with $ \lambda_p \equiv 1 $ and concave $g_\alpha (\cdot),$  converges to $x.$

(ii) If the RIC of $A$ satisfies (\ref{4D}), then the sequence $ \{x^{(p)}: p \geq 1\}, $ generated by  NTP or  NTP$_q $ with   $ \lambda_p \equiv 1$ and concave $g_\alpha (\cdot),$  converges to $x.$

\end{Cor}

The main results established in this section indicate that the significant information of the signal can be recovered by the NT-type algorithms under RIP and concavity of $ g_\alpha(\cdot).$  Corollary \ref{Cor37} further claims that under the same conditions, the $ k$-sparse signal can be exactly recovered when the measurements are accurate.

The above results also imply that the NT-type algorithms are stable for signal recovery. To establish such a stability result, let us first state the next lemma which follows immediately  from Lemma 6.23 in \cite{FR13}.

\begin{Lem}\label{Lemff}  \cite{FR13}
 Let $ k $ be an even number. Suppose that  $A \in \mathbb{R}^{m\times n} $ has the restricted isometry constant $\delta_{k } < 1.$  Given
$ \tau>0,  \xi \geq 0$ and $ \nu \in \mathbb{R}^m,$   if two vectors $x, x' \in \mathbb{R}^n $ satisfy that  $\|x'\|_0  \leq k $ and
$ \|x_S- x'\|_2 \leq  \tau \|A x_{\overline{S}}+ \nu\|_2 + \xi, $
where $S=L_k(x), $  then one has
 $$ \|x-x'\|_2  \leq \frac{1+ 2\sqrt{2} \tau } {\sqrt{k/2}} \sigma_{k/2} (x)_1+ 2 \tau \|\nu\|_2 + 2 \xi,  $$
 where $$ \sigma_{k/2} (x)_1 := \min_{u} \{\|x-u\|_1: \|u\|_0\leq k/2\}.$$
\end{Lem}

In fact, by setting $ \kappa =2, T=S:=L_k(x)$ and considering  the $ \ell_2$-norm bound,   Lemma 6.23 in \cite{FR13} immediately reduces to the above lemma. The following stability result for NT-type algorithms follows from Theorems \ref{Thm01}, \ref{Thm02} and Lemma \ref{Lemff}.

\begin{Thm} Let $ y:= Ax+\nu$ be the measurements of the signal  $x,$ where $ \nu $ are the measurement errors. Then the following two statements are valid:

(i) Suppose that the RIC of $ A$ satisfies (\ref{4A}) and $k$ is an even integer number. Then the sequence $ \{x^{(p)}: p\geq 1\},$ generated by NT or NT$_q$ with  $ \lambda_p \equiv 1 $ and concave $g_\alpha (\cdot),$  satisfies that
 \begin{align}   \|x-x^{(p)}\|_2
  & \leq    \frac{1+2\sqrt{2}C/(1-\eta)}{\sqrt{k/2}}\sigma_{k/2} (x)_1 + \frac{2C}{1-\eta} \|\nu\|_2  \nonumber  \\
& ~~~ + 2\eta^p(\|x^{(0)}\|_2+\|x\|_2), \label{3311}
\end{align}
where $C = C_1\sigma_{\max} (A)+ C_2 $, $ \sigma_{\max} (A)$ is the largest singular value of $A,$ and the constants $ C_1, C_2,\eta $ are given in Theorem \ref{Thm01}.

(ii) Suppose that the RIC of $ A$ satisfies (\ref{4D}) and  $k$ is an even integer number. Then the sequence $ \{x^{(p)}: p\geq 1\},$ generated by NTP or NTP$_q$ with    $ \lambda_p \equiv 1$ and concave $g_\alpha (\cdot),$   satisfies that
 \begin{align} \label{3322}    \|x-x^{(p)}\|_2   & \leq    \frac{1+2\sqrt{2}\widetilde{C}/(1-\rho)}{\sqrt{k/2}}\sigma_{k/2} (x)_1 + \frac{2\widetilde{C} }{1-\rho} \|\nu\|_2  \nonumber  \\
&  ~~~ + 2\rho^p(\|x^{(0)}\|_2+\|x\|_2),
\end{align}
where $\widetilde{C} = \widetilde{C}_1\sigma_{\max} (A)+ \widetilde{C}_2 $ and the constants $ \widetilde{C}_1, \widetilde{C}_2, \rho$  are given in Theorem \ref{Thm02}.
\end{Thm}

\emph{Proof.} (i) Under (\ref{4A}), Theorem \ref{Thm01} claims that the sequence $\{x^{(p)}: p\geq 1\}, $ generated by NT and NT$_q$ with    $ \lambda_p \equiv 1 $ and concave $g_\alpha (\cdot),$  satisfies (\ref{EERRBB}).  Note that $ \|A^T \nu'\|_2 \leq \sigma_{\max}(A) \|\nu'\|_2, $ where  $ \sigma_{\max}(A) $ is the largest singular value of $A.$   Thus (\ref{EERRBB}) implies that
$$  \| x^{(p+1)}-x_S \|_2 \leq \eta \|x^{(p)}-x_S \|_2 +  C \|\nu'\|_2, $$
where $ C= C_1\sigma_{\max}(A) + C_2 $ and $ \nu' = Ax_{\overline{S}}+\nu,$
and hence
\begin{align*} \|x^{(p)}-x_S\|_2 & \leq  \eta^p \|x^{0}-x_S\|_2 + C \left(\sum_{i=0}^{p-1} \eta^i\right) \|  \nu'\|_2\\
& \leq \frac{C}{1-\eta} \| \nu'\|_2 + \eta^p (\|x^{0}\|_2+ \|x \|_2),
\end{align*}
where the last inequality follows from the fact $ \sum_{i=0}^{p-1} \eta^i < 1/(1-\eta)$ and $\|x^{0}-x_S\|_2\leq \|x^{0}\|+ \|x\|_2. $
By Lemma \ref{Lemff} with $ x'=x^{(p)}$ and $ \tau =\frac{C}{1-\eta}$ and $\xi= \eta^p (\|x^{0}\|_2+ \|x \|_2)$,  we immediately obtain the bound (\ref{3311}).

(ii) By an analysis similar to Item (i) above, we can also show the bound (\ref{3322}). In fact, under   (\ref{4D}), it follows from   (\ref{4E}) in Theorem \ref{Thm02} that
$$  \| x^{(p+1)}-x_S \|_2 \leq   \rho  \|x^{(p)}-x_S \|_2 + \widetilde{C} \|\nu'\|_2, $$
where $ \widetilde{C}= \widetilde{C}_1\sigma_{\max}(A) + \widetilde{C}_2 $ and $ \nu' = Ax_{\overline{S}}+\nu.$
The inequality above implies that
 \begin{align*} \|x^{(p)}-x_S\|_2 \leq \frac{\widetilde{C}}{1-\rho} \|  \nu'\|_2 + \rho^p (\|x^{0}\|_2+ \|x \|_2),
\end{align*}
which together Lemma \ref{Lemff}  implies the bound (\ref{3322}). \hfill $ \Box$

If $ x$ is $t$-sparse (or $t$-compressible) where $ t\leq k/2,$ then $ \sigma_{k/2} (x)_1 \equiv 0 $ (or $ \sigma_{k/2} (x)_1  \approx 0$),  (\ref{3311}) and (\ref{3322}) imply that the signal $ x$ can be stably recovered by the algorithms provided that the measurements are accurate enough and the algorithms  perform  enough iterations.

\section{Further discussion on regularization and parameter} \label{sect4}

Clearly, the choice of regularization $\phi$ and parameter $\alpha$ will directly affect the numerical behavior of the algorithm.  The NT-type algorithms are developed from the first-order approximation of the function $ g_\alpha (w).$  To ensure the quality of such an approximation, the gap between $ g_\alpha (w)$ and its first-order approximation over the   set $P$ should be made as small as possible.
Thus we should choose  $\phi$ and  $\alpha$  such that the second-order term in the Taylor expansion of $ g_\alpha (w)$ can be dominated as possible by its first-order term.   One of the ideas is to strengthen the first-order approximation by suppressing the second-order term through controlling the eigenvalues of the Hessian matrix of $ g_\alpha (w)$. This might be partially achieved by properly selecting   $ \alpha.$  Moreover, the concavity of $ g_\alpha (w)$ can also be maintained by the proper selection of $\alpha,$   as shown by the next proposition.

\begin{Prop}\label{PRO01}  (i) Using the function (\ref{phi}), $g_\alpha (w) $ is concave for any given $ \alpha \geq  \alpha^*: = \lambda_{\max} (U^{(p)}A^TA U^{(p)}). $

  (ii) Using the function   (\ref{RG02}), $g_\alpha (w) $ is concave for any given parameter  $ \alpha \geq \alpha^*:= 2 \lambda_{\max} (U^{(p)}A^TA U^{(p)}) . $

  (iii) Using the function  (\ref{RG03}), $g_\alpha (w) $ is concave for any given parameter  $ \alpha \geq \alpha^*:= 4 \lambda_{\max} (U^{(p)}A^TA U^{(p)}) . $
\end{Prop}

 The proof of the proposition is given in {\sc  Appendix C}.   Note that the functions  (\ref{phi}), (\ref{RG02}) and (\ref{RG03}) are independent of $ u^{(p)}. $ It should be pointed out that all regularization functions given in Section \ref{sect2} can also be modified to include
   the vector $ u^{(p)}.$  For instance,
 the function (\ref{phi}) can be modified to involve the vector $u^{(p)} $ as follows:
\begin{align} \label{phi-new} \phi (w) &  = \sum_{i=1}^n  (u^{(p)}_i)^2 (w_i + \frac{1}{2}) (\frac{3}{2}-w_i) \nonumber  \\
 &  = (w+\frac{1}{2} \textrm{\textbf{e}})^T (U^{(p)})^2 (\frac{3}{2} \textrm{\textbf{e}}-w),  \end{align}
where $ U^{(p)}= \textrm{diag}  (u^{(p)}).$ When $ u^{(p)}_i \not=0$ for $ i=1, \dots, n$,
this function is positive over an open neighborhood of $D,$ and it remains a binary regularization specified by Definition \ref{Def01}. We also have the following observation.

\begin{Prop} \label{Pro02}  When all entries of $u^{(p)}$ are nonzero, if  $\phi$ is given by (\ref{phi-new}), then $ g_\alpha (w) $  is concave over an open neighborhood of   $D $ for any given $ \alpha \geq \alpha^*: = \lambda_{max} (A^TA).$

\end{Prop}

 \emph{Proof.} The Hessian  of $ g_\alpha (w) $  is given as
$$ \nabla^2 g_\alpha (w) = 2 U^{(p)} (A^T A - \alpha I)  U^{(p)} . $$  The result is obvious since $\nabla^2 g_\alpha (w) \preceq 0$   if and only if  $ \alpha \geq \alpha^*:  = \lambda_{max} (A^TA).$ \hfill $ \Box $

 While the concavity of $ g_\alpha (w)$ is assumed for the analysis of algorithms in Section \ref{sect3}, this assumption is largely for the convenience of analysis and might not necessarily be needed if we adopt another way of analysis. In fact, the key idea behind the derivation of the NT algorithms is the first-order approximation of the  function $ g_\alpha (w)$. Such a derivation may not require the concavity of $ g_\alpha (w).$ Even if $ g_\alpha(w) $ is not concave,  the first-order approximation of $ g_\alpha (w) $ still yields the same version of the NT-type algorithms. Also, the concavity of $g_\alpha(w) $ may not be required from a numerical point of view. The value of $ \alpha$ may be simply prescribed beforehand or be updated in the course of iterations according to some rule.

The NT-type algorithms only involve the product between matrices and vectors,  sorting and orthogonal projections, and hence the computational cost is very similar to that of HTP, SP, CoSaMP and OMP.  The computational cost of these algorithms  is much lower than   convex optimization methods   \cite{CDS98, CT05, CWB08,ZL12} and  optimal $k$-thresholding methods \cite{Z20, ZL21}. The empirical results in next section indicate that by approximate choices of the steplength $\lambda_p$ and parameter $ \alpha,$  the NT-type algorithms can  efficiently reconstruct sparse signals.

\section{Numerical Validation} \label{sect5}

  Based on random problem instances of signal recovery, we provide some preliminary experiment results to demonstrate the performance of the NT-type algorithms. For the given size and class of random matrices, we first use simulations to find suitable steplength $\lambda_p$ and parameter $\alpha$ for the algorithms.  Then we compare the performance of NT-type algorithms and SP, CoSaMP, OMP and HTP. The dimension of the synthetic sparse signal is 8000, and  the size of random matrices used in our experiments is $1000\times 8000 .$   All matrices are Gaussian matrices with columns being normalized, whose entries are independently and identically distributed (iid) random variables following the standard normal distribution. The nonzeros of  sparse signals are also assumed to be iid random variables following the standard normal distribution, and their positions are uniformly distributed. Unless otherwise specified, we use the following stopping criterion for signal recovery:
\begin{equation} \label{cri} \|x^{(p)}-x\|_2/\|x\|_2 \leq 10^{-5} , \end{equation}
where $ x$ denotes the target signal to recover and $x^{(p)}$ denotes the iterate produced by an algorithm. When $ x^{(p)}$ satisfies (\ref{cri}), the recovery of $x$  is counted as \emph{success}, otherwise \emph{unsuccess}. All results demonstrated in this section are obtained by using the function  (\ref{phi-new}) and adopting the initial point $ x^{(0)}=0$ for algorithms.  The maximum number of iterations for iterative algorithms is set to be 150. If an algorithm cannot meet the stopping criterion (\ref{cri}) in 150 iterations,  the algorithm  is then treated as  \emph{unsuccess} for the underlying recovery problem. Since NTP and NTP$_q$ use the orthogonal projection at every step, they usually outperform NT and NT$_q$ in signal recovery. Thus we only demonstrate the performance of NTP and NTP$_q$ in this section.

\subsection{Choice of regularization parameter: $\alpha$}

\begin{figure} [htp]
\includegraphics [width=0.45\textwidth,
totalheight=0.25 \textheight] {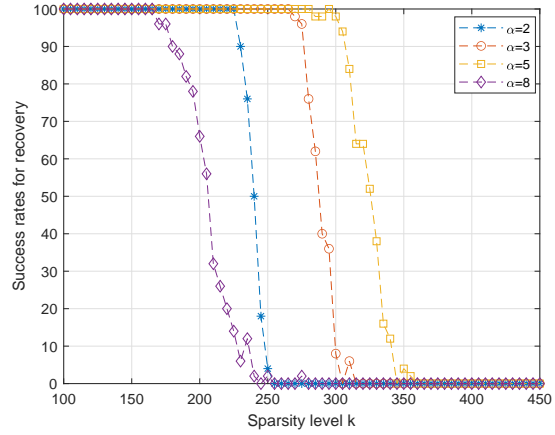}
 \caption{Success rates of NTP with  $ \alpha=2, 3, 5, 8.$}
 \label{fig-alpha}
\end{figure}

The parameter $\alpha $ in NT-type algorithms can  take any value in $(0, \infty).$   To see how the value of $ \alpha$ affects the performance of the algorithms, we compare the success frequencies of the algorithms for signal recovery with several different values of $ \alpha.$    The sparsity of  signals are ranging from 100 to 450 with stepsize 5, and the measurements are taken as $ y:=Ax$  for every realized random pair $(A,x).$ Our initial simulations indicate that the number  of inner iterations of  NTP$_q $ does not remarkably affect its overall performance. Thus we mainly consider the algorithm with $q=1,$ i.e.,  the NTP algorithm, in this experiments. For each given sparsity level $k$, 100 random trials are used to estimate the success rates of the  NTP, and the algorithm is performed up to a total of 150 iterations. Fig. \ref{fig-alpha} shows the results for NTP with several different values of $ \alpha.$   The steplength is fixed as $ \lambda_p \equiv 2$ (this stepsize is suggested from simulations. See the next subsection for details).
The results in Fig.\ref{fig-alpha} indicates that for the Gaussian matrices of size $1000\times 8000, $  $ \alpha=5$ is a reasonable choice for NTP.
However, we should keep in mind that such a choice of the parameter depending on the size/class of problem instances should not be generalized to  other size/class of the problems.

\subsection{Influence of iterative steplength: $ \lambda_p$}

\begin{figure} [htp]
\includegraphics [width=0.45\textwidth,
totalheight=0.25 \textheight] {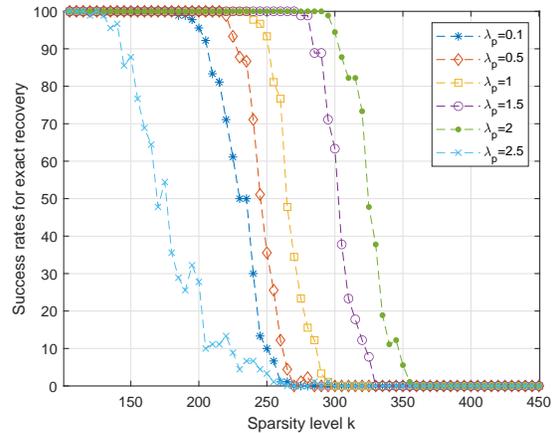}
 \caption{Success rates of NTP with different $\lambda_p$}
 \label{fig-lambda}
\end{figure}

In traditional function minimization scenarios, receding in the direction of negative gradient of a function from the current iterate by a small steplength can guarantee the descent of the function. Thus the steplength in traditional optimization method is usually smaller than 1. However, when solving the SCO problem (\ref{L0}), this property of steplength is lost since  the main purpose of the iterative algorithm for SCO is to identify the correct support of the solution instead of seeking immediate descent of the objective value.  Thus when solving the SCO, the steplength of an iterative algorithm may not necessarily be smaller than 1. Indeed, a large amount of simulations indicate that the efficient steplength longer than 1 is commonly observed when   solving a SCO problem. For $ 1000 \times 8000$ Gaussian sensing matrices, we carried out experiments to show how the steplength might affect the performance of NTP.  The random pair of $ (A,x)$   are generated exactly as in previous subsection, and we set $ \alpha=5 $ for NTP in our experiments.     For every given sparsity level and steplength, 100 trials were used to estimate the success rate  of the algorithm.  The results for   $ \lambda_p=0.1, 0.5, 1, 1.5, 2$ and $2.5 $ are summarized in Fig. \ref{fig-lambda} which indicates that $\lambda_p =2$ is relatively good among the ones being tested.

\subsection{Number of iterations}
\begin{figure} [htp]
\includegraphics [width=0.45\textwidth,
totalheight=0.25 \textheight] {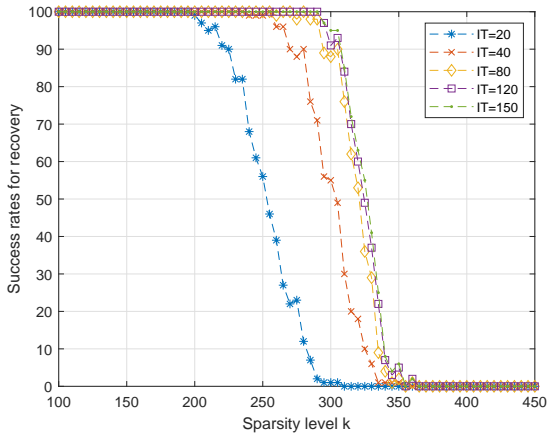}
 \caption{Comparison of success rates of NTP which is performed  different number of iterations. }
 \label{fig-iteration}
\end{figure}

 The random instances of SCO  problems are generated in the same way as previous subsections. As suggested by the  above experiments, we set $ \alpha \equiv5$ and $ \lambda_p \equiv 2$   in NTP.  We test the recovery ability of NTP by performing the algorithm with several different number of iterations: 20, 40, 80, 120 and 150.  For every sparsity level $k$ ranging from 100 to 450 with stepsize 5, we use 100 random trials to estimate the success rate of the algorithm.  The results are summarized in Fig. \ref{fig-iteration} in which IT means the total number of iterations performed. The result indicates that the overall performance of NTP becomes better and better as the total number of  iterations   increases.  However, after the algorithm is performed enough iterations, say, $\textrm{IT} > 100,$ any further iterations   would not remarkably improve its performance.   This experiment indicates that for the class of SCO problems being tested,  150 iterations is generally  enough to ensure the algorithm to work well for solving such a class of problems. Thus in the remaining experiments, we set 150 as the maximum number of iterations, and if the algorithm cannot solve the problem  in 150 iterations, the algorithm is treated as \emph{unsuccess} for the problem.

\subsection{Runtime Comparison}

\begin{figure} [htp]
\includegraphics [width=0.45\textwidth,
totalheight=0.25 \textheight] {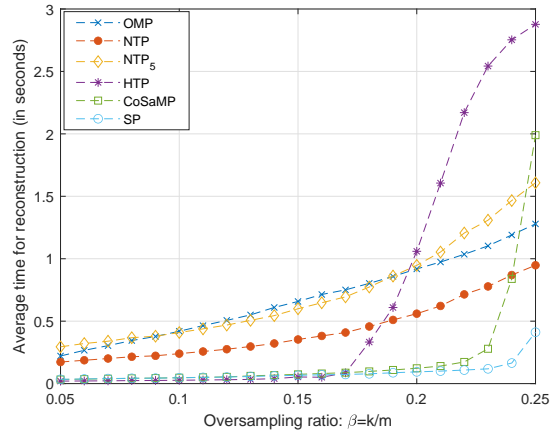}
 \caption{Runtime comparison of several algorithms}
 \label{fig-time}
\end{figure}

Given $m=1000$ and $ n= 8000,$ we consider a series of vectors with different sparsity levels:   $k = \beta m$ where $ \beta$ (called the oversampling ratio) takes 11 different values between 0.05 and 0.25. Namely, $ \beta = 0.05+0.2i $ for $i=0, 1, \dots, 10.$ For every given  $ k= \beta m$, we  use NTP, NTP$_5$,   OMP, HTP, CoSaMP and  SP  to recover the $k$-sparse vector through the  measurements   $y: = Ax.$ For every given $ \beta$, the average time required to recover the sparse vector is calculated based on 50 random trials. The results are shown in Fig. \ref{fig-time}, from which it can be seen that  the average time required by the NTP algorithm is lower than that of OMP and NTP$_5$, and the average time required by NTP$_5$ to recover the signals is very similar to that of OMP.  We also observe that when $ \beta$ is relatively small,  the runtime of  traditional  CoSaMP, SP and HTP is lower than that of NTP and  NTP$_5. $

\subsection{Comparison of success frequencies}

\begin{figure} [htp]
\includegraphics [width=0.45\textwidth,
totalheight=0.25 \textheight] {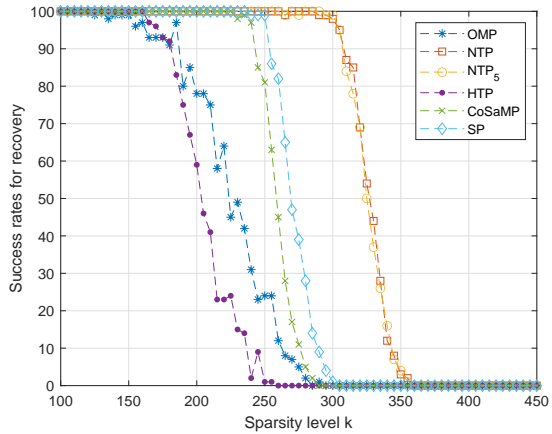}
 \caption{Comparison of success frequencies in noiseless situations. 100 random examples were attempted and the recovery criterion (\ref{cri}) was adopted. }
 \label{fig-accurate}
\end{figure}

The performance of NTP, NTP$_q$ and HTP depends on the choice of steplength $\lambda_p. $ The performance of NTP and NTP$_q$ also depends on the choice of $\alpha. $ We set $ \alpha=5$ and $ \lambda_p \equiv 2$ in NTP and NTP$_5,$  and we use the same steplength $ \lambda_p \equiv 2$ in  HTP.   For every random pair  of $ (A, x),$ where $ x$ is a sparse vector, the accurate measurements are given by $ y: = Ax.$   The success frequencies of   NTP, NTP$_5$,  HTP, OMP, SP and CoSaMP  are estimated through 100 random examples of SCO problems.  All iterative algorithms are performed a total of 150 iterations except for the OMP which by its structure performs a total of $k$ iterations, where $k$ is the sparsity level of the sparse signal.
The results are demonstrated in Fig. \ref{fig-accurate}, from which one can see that NTP and NTP$_5$ perform similarly and they outperforms the traditional HTP, OMP, SP and  CoSaMP.  However,  OMP, SP, CoSaMP  have their own advantages over HTP and NT-type algorithms  in the sense that  they do not need any parameter or steplength  to solve the SCO problems.

\begin{figure} [htp]
\includegraphics [width=0.45\textwidth,
totalheight=0.25 \textheight] {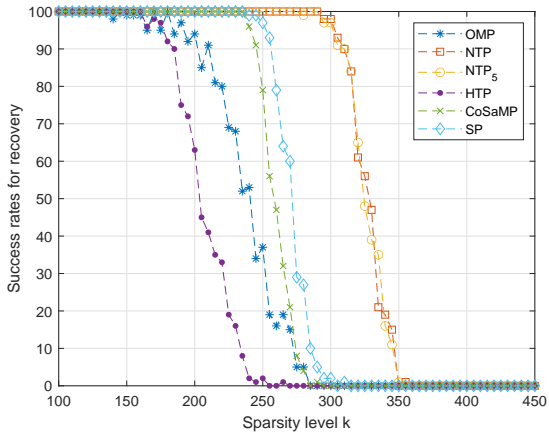}
 \caption{Comparison of success rates in noisy situations: $ \xi = 0.001.$  100 random examples were attempted and the recovery criterion (\ref{NERR}) was adopted.}
 \label{fig-inaccurate}
\end{figure}

\begin{figure} [htp]
\includegraphics [width=0.45\textwidth,
totalheight=0.25 \textheight] {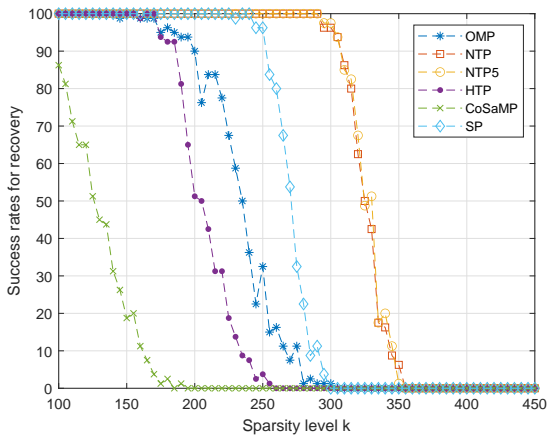}
 \caption{Comparison of success rates in noisy situations: $ \xi = 0.01.$ 100 random examples were attempted and the recovery criterion (\ref{NERR}) was adopted.}
 \label{fig-inaccurate1}
\end{figure}

We also compare the algorithms in the setting of noisy measurements. In this setting, the recovery criterion is set as \begin{equation} \label{NERR}  \|x^{(p)}-x\|_2/\|x\|_2 \leq 10^{-3} , \end{equation} and the inaccurate measurements are given as $ y:= Ax +\xi v,$ where $v$ is a normalized Gaussian noise vector and $\xi>0 $ reflects the noise level. The  success rates of algorithms  are estimated with 100 random examples, and the algorithms are performed the same number of iterations as in the noiseless case.  $\alpha=5$ and $ \lambda_p\equiv 2$ are still used in this experiment. By setting $ \xi = 0.001,$ the results are shown in Fig. \ref{fig-inaccurate} which indicates that NTP and NTP$_5$  are  still efficient and can compete with several other algorithms in sparse signal recovery.   Changing the noise level from $ \xi = 0.001$ to $ \xi = 0.01$ and repeating the experiment, we obtain the results demonstrated in Fig. \ref{fig-inaccurate1} which indicate that such a chance in noise level affects significantly the performance of CoSaMP, but does not remarkably affect the performance of several other algorithms. From Fig. \ref {fig-inaccurate} and Fig. \ref{fig-inaccurate1}, it can be observed that NTP and NTP$_5$ are robust in signal recovery when measurements are slightly inaccurate.

\section{Conclusions}

The natural thresholding (NT) algorithms for sparse signal recovery have been developed in this paper. This class of algorithms was derived through the first-order approximation of the regularized optimization model for optimal $k$-thresholding. The concept of concave binary regularization plays a vital role in this development. The guaranteed performance of the NT-type algorithms in sparse signal recovery has been shown under the restricted isometry property together with the concavity of the objective function of the regularized optimization model. Simulations via random examples show that the proposed algorithms can compete with the traditional HTP, OMP, CoSaMP and SP algorithms.

The concavity is imposed to show the main results (Theorems \ref{Thm01} and \ref{Thm02}) in this paper. A worthwhile future work might be  the analysis of NT-type algorithms under other nonconvex conditions. Another future work might be the development of an algorithm with adaptive steplength $\lambda_p$ and parameter $ \alpha $  which are updated during iterations according to a certain rule, so that the efficiency and performance of the NT-type algorithms would be further enhanced.

\section*{Appendix A: Proof of Lemma \ref{Lemma-penality}}

 \emph{Proof.} Let $ \phi(w) $ be a binary regularization function (see Definition \ref{Def01}).  Let $ w(\alpha) $ denote a solution to the problem (\ref{HT-QP-Relax-2}) with parameter $ \alpha>0.$   Let $ \{\alpha_{\ell}\} $ be any positive sequence which is strictly increasing  and tends to $ \infty$ as $ \ell\to \infty.$  For notational convenience, denote by $ w_{\ell}:= w(\alpha_{\ell}) $   the optimal solution  to (\ref{HT-QP-Relax-2}) with $ \alpha:=\alpha_{\ell} .$    Thus by the optimality of $w_{\ell}$ and $w_{\ell-1} $, one has $$ g_{\alpha_{\ell}} ( w_{\ell}) \leq g_{\alpha_{\ell}} ( w_{\ell-1}),
 ~~  g_{\alpha_{\ell-1}} ( w_{\ell-1}) \leq g_{\alpha_{\ell-1}} (w_{\ell}) $$ That is,
 $$ f( w_{\ell}) + \alpha_{\ell} \phi (w_{\ell} ) \leq f( w_{\ell-1}) + \alpha_{\ell} \phi (w_{\ell-1}), $$
\begin{equation} \label{332211}  f( w_{\ell-1}) + \alpha_{\ell-1} \phi (w_{\ell-1}) \leq f( w_{\ell}) + \alpha_{\ell -1} \phi (w_{\ell}).  \end{equation}
  Note that  $ \alpha_{\ell-1}< \alpha_{\ell}.$ Adding the two inequalities, cancelling and combining terms yields
   $$   \phi (w_{\ell} ) \leq   \phi (w_{\ell-1}), $$
  i.e., the sequence $\{ \phi (w_{\ell} ) \}$ is nonincreasing. Thus $ \phi (w_{\ell} ) \to \phi^*$ as $ \ell \to \infty,$ where $ \phi^*$ is a nonnegative number since the nonnegativeness of $\phi(w) $ over the set $P \subseteq D .$ Denote by $ \phi_{\min}$ the minimum of $ \phi(w)$ over $P.$ By the definition of binary regularization, $\phi_{\min} $ attains and only attains at any $k$-sparse binary vector in $P.$ It immediately follows from $ \phi(w_{\ell}) \geq \phi_{\min}$ that $\phi^* \geq \phi_{\min}. $ We now further show that $ \phi^*= \phi_{\min}. $
  Let $ w^*$ be any given $k$-sparse binary vector in $P.$ Then by the optimality of $w_{\ell}$ again, we have
  $$ f(w_{\ell}) + \alpha_{\ell} \phi (w_{\ell} ) \leq f(w^*) +  \alpha_{\ell} \phi(w^*) = f(w^*) +  \alpha_{\ell} \phi_{\min}   $$
  which together with the fact $f(w_{\ell})\geq 0 $ implies that
 $$ \alpha_{\ell} (\phi (w_{\ell} )- \phi_{\min} )\leq f(w^*)  $$
  Therefore, one has
  $$ 0\leq \phi^*-\phi_{\min} \leq \phi (w_{\ell} ) -  \phi_{\min} \leq f(w^*)/\alpha_{\ell},$$
   which together with $ \alpha_{\ell}\to \infty$   implies that  $\phi^* =  \phi_{\min}.$
  This together with the convergence of $ \{ \phi (w_{\ell})\} $ and continuity of $ \phi$  implies that any accumulation point of $ w_{\ell}$  must be a $k$-sparse and binary vector (since the minimum of $ \phi$ over $P$ attains only at $k$-sparse binary vector). We now further show that the sequence  $\{ f(w_{\ell})\}$ is nondecreasing and convergent to the optimal value of (\ref{conrel}) and that any accumulation point of $ \{w_{\ell}\} $ is an optimal solution to (\ref{conrel}). In fact, by (\ref{332211}), we have
  $$  f( w_{\ell-1}) + \alpha_{\ell-1} \phi (w_{\ell-1}) \leq f( w_{\ell}) + \alpha_{\ell -1} \phi (w_{\ell}), $$
  which implies that  $$ f( w_{\ell-1}) -f( w_{\ell}) \leq \alpha_{\ell-1} ( \phi (w_{\ell}) - \phi (w_{\ell-1})) \leq 0$$
  thus $f( w_{\ell-1})  \leq f( w_{\ell}) $ which means the sequence $ \{ f(w_{\ell})\}$ is nondecreasing.  Assume that $ \widehat{w}$ is any optimal solution of (\ref{conrel}), by the optimality of $ w_{\ell-1},$ one has
  $$ f( w_{\ell-1}) + \alpha_{\ell-1} \phi (w_{\ell-1}) \leq f( \widehat{w}) + \alpha_{\ell -1} \phi (\widehat{w}), $$
  where $ \phi (\widehat{w})$ must be the minimum value of $ \phi$ over $D$ since $ \widehat{w}$ is $k$-sparse and binary.
  Thus the above inequality implies that
  $$ f( w_{\ell-1}) \leq  f(\widehat{w} ) + \alpha_{\ell-1} ( \phi (\widehat{w}) - \phi (w_{\ell-1}))  \leq f(\widehat{w}), $$
  so the nondecreasing sequence $ \{ f(w_{\ell})\}$ is bounded above, and hence $$ f(w_{\ell}) \to f^* \leq f(\widehat{w})  \textrm{ as }    k \to \infty .$$
   Let $ \overline{w} $ be any accumulation of $ w_{\ell}$ (as shown above, $ \overline{w} $ is $k$-sparse and binary). It follows from the above relation that $f^* = f(\overline{w}) \leq f(\widehat{w}).$   Since the right-hand side is the minimum value of (\ref{conrel}). Thus we deduce that  $f(\overline{w})= f(\widehat{w} )$. So any accumulation point of $\{w_{\ell}\}$ is the optimal solution to (\ref{conrel}).  \hfill    $\Box$

\section*{Appendix B: Proof of Theorem \ref{Thm01}}
\emph{Proof}:  Let $x^{(p)}$ be the current iterate generated by NT or NT$_q ,$ and let $u^{(p)}$ be given by (\ref{uupp}).
  The inner iterations of  NT and NT$_q$ start  from the vector $ w^- $  which satisfies  $ w^- \otimes u^{(p)}= {\cal H}_k(u^{(p)}). $ The NT algorithm performs only one inner iteration, and NT$_q$   performs $q$ times of inner iterations. It follows from Lemma \ref{Lem-a} that the output $ w^+$  of the inner iterations of NT or NT$_q$ algorithm satisfies
    $$\| y- A (u^{(p)} \otimes w^+)\|_2 \leq \| y- A (u^{(p)} \otimes w^-)\|_2.$$
Note that  $  u^{(p)} \otimes w^- = {\cal H}_k(u^{(p)}) $ and   $x^{(p+1)} = u^{(p)} \otimes w^+$. The  above inequality is written as
\begin{equation} \label{IIEEQQ}  \| y-A x^{(p+1)}\|_2  \leq \|y-A {\cal H} _k(u^{(p)})\|_2. \end{equation}  We also note that
\begin{equation}\label{F33}
y= Ax+\nu = Ax_S + (Ax_{\overline{S}}+\nu) =Ax_S + \nu',
\end{equation} where $\nu'= Ax_{\overline{S}}+\nu. $
 Since $ x_S-x^{(p+1)}$ is $(2k)$-sparse, by (\ref{F33}) and the triangular inequality and the definition of $ \delta_{2k}, $  one has
\begin{align}
\| y    -A x^{(p+1)}\|_2
& = \| A(x_S-x^{(p+1)}) +   \nu'  \|_2  \nonumber \\
&\geq \| A(x_S-x^{(p+1)})\|_2 -  \|\nu' \|_2 \nonumber \\
& \geq \sqrt{1-\delta_{2k}} \|x_S-x^{(p+1)}\|_2 - \|\nu'\|_2.    \label{IIEEQQ01}
\end{align}
Similarly, since
  $x_S-{\cal H}_k(u^{(p)})$ is $(2k)$-sparse, we have
\begin{align}  \|y  &    -A {\cal H} _k(u^{(p)})\|_2  \nonumber\\
& =  \|A(x_S- {\cal H} _k(u^{(p)}))+  \nu' \|_2 \nonumber\\
 & \leq  \|A(x_S- {\cal H} _k(u^{(p)}))\|_2+ \| \nu' \|_2\nonumber \\
 & \leq \sqrt{1+\delta_{2k}} \|x_S- {\cal H} _k(u^{(p)})\|_2   + \| \nu' \|_2    \label{IIEEQQ02}
\end{align}
Merging (\ref{IIEEQQ}), (\ref{IIEEQQ01}) and  (\ref{IIEEQQ02}) yields
\begin{align} \label{RRTT}    \| x_S   &   -   x^{(p+1)}\|_2  \nonumber \\   &  \leq       \sqrt{ \frac{ 1+\delta_{2k}}{1-\delta_{2k} } }   \| x_S- {\cal H}_k(u^{(p)})\|_2  + \frac{2\| \nu' \|_2}{\sqrt{1-\delta_{2k}}} .  \end{align}
Let $ S^* = \textrm{supp} ({\cal H}_k(u^{(p)})).  $    Note that the set $ S^* \cup S $, where  $ S= L_k(x),$  can be decomposed into three disjoint sets $ S^*\backslash S, S\backslash S^* $ and $ S \cap S^*.$ Thus
\begin{align} \label{XXUU}
         \|(x_S- &   u^{(p)})_{S^* \cup S}\|_2^2      =       \|(x_S-u^{(p)})_{S^* \backslash S}\|_2^2   \nonumber \\
       &     + \|(x_S-u^{(p)})_{S \backslash S^*}\|_2^2  + \|(x_S-u^{(p)})_{S \cap S^*}\|_2^2.
\end{align}
 Without loss of generality, we assume that
 $ \|(x_S-      u^{(p)})_{S^* \cup S}\|_2^2  \not = 0.  $ Under this assumption,
    one of the three terms on the right-hand of (\ref{XXUU})
is nonzero. Without loss of generality, we assume that the first term on the right-hand side of (\ref{XXUU}) is nonzero, i.e., $\|(x_S-u^{(p)})_{S^* \backslash S}\|_2^2 \not=0.$ Then there exists numbers $ \mu_1 \geq 0 $ and $\mu_2\geq 0$ such that
\begin{align} \|(x_S-u^{(p)})_{S \backslash S^*}\|_2   & = \mu_1 \|(x_S-u^{(p)})_{S^* \backslash S}\|_2,      \label{EQ001}    \\
  \|(x_S-u^{(p)})_{S \cap S^*}\|_2  & = \mu_2 \|(x_S-u^{(p)})_{S^* \backslash S}\|_2.    \label{EQ02}
    \end{align}
If $ \|(x_S- u^{(p)})_{S^* \cup S}\|_2^2    =0, $ then all terms on the right-hand of (\ref{XXUU}) are zero, and hence (\ref{EQ001}) and (\ref{EQ02}) are still valid in this case.
Substituting (\ref{EQ001}) and (\ref{EQ02}) into (\ref{XXUU}) yields
\begin{equation}  \label{SSEE}  \|(x_S-      u^{(p)})_{S^* \cup S}\|_2^2 = (1+ \mu_1^2 + \mu_2^2) \|(x_S-u^{(p)})_{S^* \backslash S}\|_2^2.
\end{equation}
By the definition of $S$ and $ S^*$, we have
\begin{align*}    \| x_S     - & {\cal  H}_k(u^{(p)})\|_2^2
  =     \|(x_S-{\cal H}_k(u^{(p)}))_{S^* \cup S}\|_2^2 \\
  =  & \|(x_S-{\cal H}_k(u^{(p)}))_{S^* \backslash S}\|_2^2 +\|(x_S-{\cal H}_k(u^{(p)}))_{S^* \cap S}\|_2^2 \\
    &   + \|(x_S-{\cal H}_k(u^{(p)}))_{ S\backslash S^*}\|_2^2 \\
  =  &   \|(x_S-u^{(p)})_{S^* \backslash S}\|_2^2 +\|(x_S-u^{(p)})_{S^* \cap S}\|_2^2 \\
   &   + \|(x_S-{\cal H}_k(u^{(p)}))_{ S\backslash S^*}\|_2^2.
   \end{align*}
Thus by  Lemma \ref{Lem4433}, we have
\begin{align*}    \| x_S     - & {\cal  H}_k(u^{(p)})\|_2^2 \\
  \leq  &  \|(x_S-u^{(p)})_{S^* \backslash S}\|_2^2 +\|(x_S-u^{(p)})_{S^* \cap S}\|_2^2 \\
 &   +  \left( \|(x_S-u^{(p)})_{S^* \backslash S}\|_2  +  \|(x_S-u^{(p)})_{S\backslash S^*  }\|_2\right) ^2  \\
  =  &   \|(x_S-u^{(p)})_{S^* \backslash S}\|_2^2 +\|(x_S-u^{(p)})_{S^* \cap S}\|_2^2 \\
&   + \|(x_S-u^{(p)})_{S^*\backslash S}\|_2^2 +   \|(x_S-u^{(p)})_{S \backslash  S^*  }\|_2^2    \\
 & + 2  \|(x_S-u^{(p)})_{S^*\backslash S}\|_2  \|(x_S-u^{(p)})_{S  \backslash S ^* }\|_2 \\
  =  &   \|(x_S-u^{(p)})_{S^* \cup S}\|_2^2 +  \|(x_S-u^{(p)})_{S^*\backslash S}\|_2^2 \\
  & + 2  \|(x_S-u^{(p)})_{S^*\backslash S}\|_2  \|(x_S-u^{(p)})_{S  \backslash S ^* }\|_2 .
\end{align*}
By (\ref{EQ001}) and (\ref{SSEE}), the inequality above can be written as
\begin{align}  &  \| x_S     -   {\cal H}_k(u^{(p)})\|_2^2 \nonumber\\
  & \leq      \|(x_S-u^{(p)})_{S^* \cup S}\|_2^2 +  (1+2\mu_1) \|(x_S-u^{(p)})_{S^*\backslash S}\|_2^2 \nonumber\\
   & =   \|(x_S-u^{(p)})_{S^* \cup S}\|_2^2 + \frac{1+2\mu_1}{ 1+\mu_1^2+\mu_2^2} \|(x_S-u^{(p)})_{S^* \cup S}\|_2^2 \nonumber\\
  & =  \frac{(1+\mu_1)^2 +1+ \mu_2^2}{ 1+\mu_1^2+\mu_2^2}  \|(x_S-u^{(p)})_{S^* \cup S}\|_2^2\nonumber \\
 &  \leq     \max_{\mu_1\geq 0}  \left(\max_{ \mu_2\geq 0} \frac{(1+\mu_1)^2 + 1+  \mu_2^2}{ 1+\mu_1^2+\mu_2^2} \right)   \|(x_S-u^{(p)})_{S^* \cup S}\|_2^2\nonumber\\
  & =   \mu^* \|(x_S-u^{(p)})_{S^* \cup S}\|_2^2,    \label{ER01}
\end{align}
where \begin{align*}  \mu^*  & =    \max_{\mu_1\geq 0}  \left(\max_{ \mu_2\geq 0} \frac{(1+\mu_1)^2  +1+ \mu_2^2}{ 1+\mu_1^2+\mu_2^2} \right) \\
 & =  \max_{\mu_1\geq 0}    \frac{(1+\mu_1)^2 +1 }{ 1+\mu_1^2}   = \frac{5+\sqrt{5}}{5-\sqrt{5}} =\left( \frac{\sqrt{5}+1}{2}\right)^2 .
\end{align*}
The maximum above with respect to $ \mu_2$ is achieved at $ \mu_2=0,$ and the maximum with respect to $ \mu_1 $ is achieved at $ \mu_1 = (\sqrt{5}-1)/2.$
By (\ref{F33}) and Lemma \ref{333111},   we obtain that
\begin{align} &  \|(x_S-u^{(p)})_{S^* \cup S}\|_2 \nonumber  \\
& =\|(x_S-x^{(p)} -A^T (y-A x^{(p)}))_{S^* \cup S}\|_2  \nonumber \\
 & = \|[(I-A^TA)(x_S-x^{(p)})]_{S^* \cup S} + [ A^T \nu'] _{S^* \cup S}\|_2   \nonumber \\
& \leq \|[ (I-A^TA)(x_S-x^{(p)})]_{S^* \cup S}\|_2   + \| A^T \nu'\|_2  \nonumber \\
&  \leq \delta_{3k}  \|x_S-x^{(p)}\|_2 + \| A^T  \nu' \|_2 .   \label{ER02}
\end{align}
 The last inequality follows from Lemma \ref{333111} with $|(S^* \cup S)\cup \textrm{supp}(x_S-x^p)|\leq 3k.$
Combining   (\ref{ER01}) and (\ref{ER02}) yields
\begin{align*} \| x_S- {\cal H} _k(u^{(p)})\|_2  & \leq \frac{\sqrt{5}+1}{2} (\delta_{3k} \|x_S-x^{(p)}\|_2  + \| A^T\nu'\|_2 ).
 \end{align*}
Denote by $ \omega  = (\sqrt{5}+1)/2. $  Merging the above relation with (\ref{RRTT}), we obtain
\begin{align*}   \| x_S       -x^{(p+1)}\|_2
  &  \leq      \omega  \delta _{3k}\sqrt{ \frac{ 1+\delta_{2k}  }  {1-\delta_{2k} } } \|x_S-x^{(p)}\|_2   \\
 &  ~~ +  \omega\|A^T\nu'\|_2\sqrt{ \frac{ 1+\delta_{2k}  }  {1-\delta_{2k} } } +\frac{2\|\nu'\|_2}{\sqrt{1-\delta_{2k}}}  \\
   &  = \eta \|x_S-x^{(p)}\|_2 + C_1 \|A^T\nu'\|_2+ C_2\|\nu'\|_2,
\end{align*}
where $ C_1,$  $ C_2$ and $ \eta $ are given by (\ref{C1C2}) and (\ref{RE}), respectively.
 Note that $ \delta_{2k} \leq \delta_{3k}$ which implies that $ \sqrt{ \frac{ 1+\delta_{2k}  }  {1-\delta_{2k} } } \leq \sqrt{ \frac{ 1+\delta_{3k}  }  {1-\delta_{3k} } }. $ Thus to ensure $ \eta<1,$ it is sufficient to require that $  \omega \delta_{3k}   \sqrt{   \frac{ 1+\delta_{3k}}{1-\delta_{3k}}  }  <1 $ which,
  by squaring and rearranging terms, is written equivalently as
 $$ h(\delta_{3k}):=\omega \delta_{3k} ^3 + \omega \delta_{3k} ^2 + \delta_{3k} -1<0.$$  It is very easy to check that
 the inequality above can be guaranteed if $ \delta_{3k} < \gamma^* \approx  0.4712 ,$ where $ \gamma^*$ is the unique real root of the univariate equation $  h(t)= \omega t^3+\omega t^2 +t-1=0 $ in the interval [0,1].  In fact, $ h(t) $ is strictly increasing over $[0,1],$ and $ h(0) =-1<0 $. This implies that  $ h(t) < 0$ for any $ t < \gamma^*. $  Thus $ \eta<1$ is guaranteed if $ \delta_{3k} <\gamma^*. $    \hfill  $ \Box $

\section*{Appendix C: Proof of Proposition \ref{PRO01} }

\emph{Proof.} (i) This has been shown in section \ref{sect2}.C.

(ii) Consider the regularization (\ref{RG02}). It is easy to verify that the second-order derivative of  $\log (1+ (w_i+1/2) (3/2-w_i) ) $  with respect to $w_i$ is given by   $   \frac{-6+2\beta_i}{(1+\beta_i)^2},$ where $ \beta_i= (w_i+1/2) (3/2-w_i).$ Clearly, $ 0<\beta_i\leq 1$ over the interval  $ -1/2< w_i < 3/2$ which is an open neighborhood of the closed interval   $ 0\leq w_i \leq 1.$ Thus it is easy to verify that  the supremum of $ \frac{-6+2\beta_i}{(1+\beta_i)^2}$ over the open interval $ -1/2< w_i < 3/2$  is $  -1.$  Therefore,
{\small \begin{align*}   \nabla^2 g_\alpha (w)
  & =  2 U^{(p)}A^TA U^{(p)} + \alpha  \textrm{diag} \left( \frac{-6+2\beta_i}{(1+\beta_i)^2},  i=1, \dots, n    \right) \nonumber  \\
                                     & =  2 U^{(p)}A^TA U^{(p)} -  \alpha I \nonumber \\
                                      &  ~~~ + \alpha \left[ I +   \textrm{diag} \left( \frac{-6+2\beta_i}{(1+\beta_i)^2}, i=1, \dots, n  \right)\right]  \nonumber\\
&  \preceq  2U^{(p)}A^TA U^{(p)} -   \alpha I,
\end{align*}   }
where the last relation  follows from the fact   $$  I   +  \textrm{diag}  \left(\frac{-6+2\beta_i}{(1+\beta_i)^2}, i=1, \dots, n   \right)  \preceq 0. $$ Thus if  $ \alpha \geq \alpha^*:=2 \lambda_{\max} (U^{(p)}A^TA U^{(p)}), $    then  $ \nabla^2 g_\alpha (w) \preceq 0$ and hence   the function $g_\alpha (w) $  is concave.

(iii) This can be shown by an analysis similar to (ii). The detail is omitted.  \hfill $ \Box$

\end{document}